\documentclass[fleqn,usenatbib]{mnras}

\usepackage{newtxtext,newtxmath, pdflscape}

\usepackage[T1]{fontenc}

\DeclareRobustCommand{\VAN}[3]{#2}
\let\VANthebibliography\thebibliography
\def\thebibliography{\DeclareRobustCommand{\VAN}[3]{##3}\VANthebibliography}

\usepackage{graphicx}	
\usepackage{amsmath}	
\usepackage{multirow}   
\usepackage{tabularx}
\usepackage{float}      
\usepackage{comment}    

\usepackage{threeparttable} 
\newcommand{\fullPScount}{359}  
\newcommand{\nocensoredPScount}{346}

\defcitealias{Kerrison2024}{Paper I}
\usepackage[normalem]{ulem}

\title{RadioSED II: discovering the peaked spectrum radio sources in Stripe 82}

\author[E.F. Kerrison et al.]{Emily F. Kerrison,$^{1,2,3}$\thanks{E-mail: emily.kerrison@sydney.edu.au}
Elaine M. Sadler,$^{1,2,3}$
Vanessa A. Moss$^{3,1}$
Elizabeth K. Mahony$^{3}$
\newauthor{Laura Driessen$^{1}$, 
Kathryn Ross$^{4,5}$, Kovi Rose$^{1,3}$, Dougal Dobie$^{1,6}$, 
\& Tara Murphy$^{1,6}$}
\\
$^{1}$Sydney Institute for Astronomy, School of Physics A28, University of Sydney, NSW 2006, Australia \\
$^{2}$ARC Centre of Excellence for All Sky Astrophysics in 3 Dimensions (ASTRO 3D) \\
$^{3}$ATNF, CSIRO Space and Astronomy, PO Box 76, Epping, NSW 1710, Australia \\
$^{4}$ICRAR, International Centre for Radio Astronomy Research, Curtin University, Bentley, WA 6102, Australia \\
$^{5}$AusSRC, Australian SKA Regional Centre, Curtin University, Bentley, WA, 6102, Australia \\
$^{6}$ARC Centre of Excellence for Gravitational Wave Discovery (OzGrav), Hawthorn, Victoria, 3122, Australia}

\date{Accepted XXX. Received YYY; in original form ZZZ}

\pubyear{2025}

\begin{document}
\label{firstpage}
\pagerange{\pageref{firstpage}--\pageref{lastpage}}
\maketitle

\begin{abstract}

This paper is the second in a series presenting \textsc{RadioSED}, a Bayesian inference framework for constructing, modelling and classifying radio spectral energy distributions from publicly-available surveys. We focus here on the application of our framework to SDSS Stripe 82. Not only do we recover all eleven previously-published peaked spectrum sources from the literature within this region, but we increase the number of known peaked spectrum sources here by more than an order of magnitude. We investigate the variability properties of our peaked spectrum sample, and find that overall they exhibit a low degree of variability, consistent with previous samples of peaked spectrum active galactic nuclei. The multiwavelength properties of these sources reveal that we have selected a population comprising largely distant ($z \geq 1$), powerful active galaxies. We find that the most compact jets are located preferentially in quasar-type hosts, with galaxy-type hosts home to slightly more extended radio structures. We discuss these findings in the context of current and forthcoming large area radio surveys.

\end{abstract}




\section{Introduction}

\begin{figure*}
    \centering
    \includegraphics[width=0.95\linewidth]{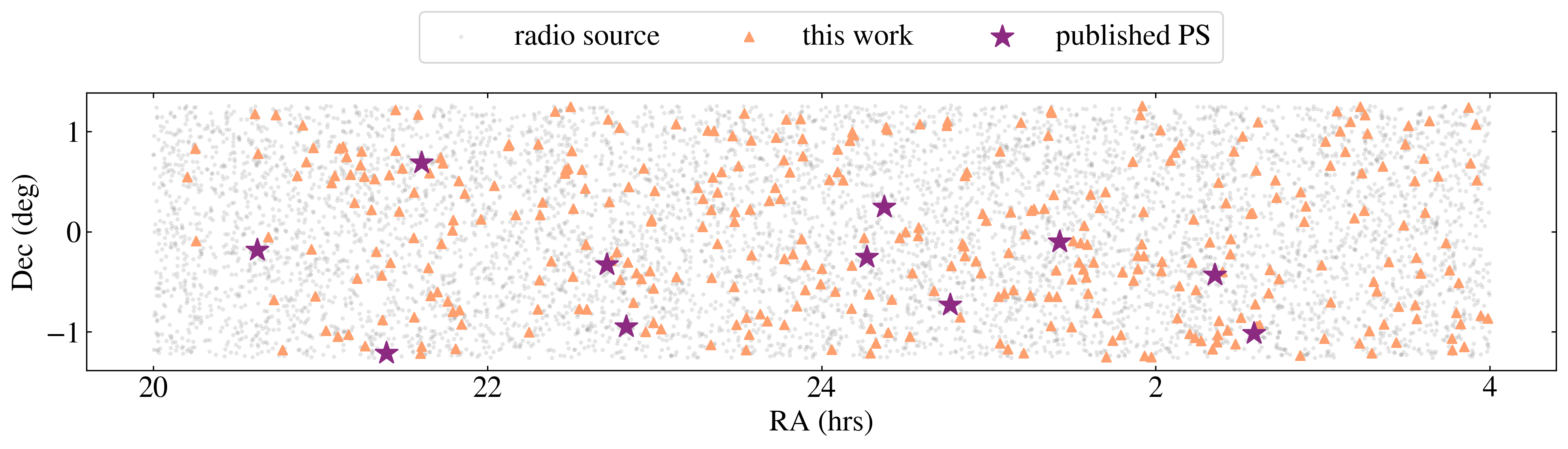}
    \caption{A sky plot of the Stripe 82 field, showing the distribution of our sample of sources with peaked SEDs (orange triangles) compared to all PS sources previously published in the literature (purple stars). All RACS-low sources in the field are overplotted as light grey points for reference.}
    \label{fig:skydist-sample}
\end{figure*}

In \cite{Kerrison2024} (henceforth \citetalias{Kerrison2024}) we presented a new Bayesian inference framework to construct and model radio spectral energy distributions called \textsc{RadioSED}. Here, we put this framework to the test by using it to identify a new sample of peaked spectrum radio active galactic nuclei (`AGNs'), the properties of which are examined using available multiwavelength data.

Young radio AGNs within a few thousand years of triggering are thought to possess radio jets which have not yet expanded out beyond the gas-rich environment of their host, with the resulting strong magnetic fields and dense ambient medium leading to absorption at low radio frequencies. It is this absorption which leads to a peak in their broadband radio spectral energy distribution (SED), and from which they commonly take the name `Peaked Spectrum' (PS) sources \citep{ODea2021}. However, this is not the only reason for a radio AGN to exhibit a peaked spectrum, indeed some PS sources are thought to be not necessarily young, but impeded in their growth by the dense, ambient medium of their host \citep[e.g.][]{Callingham2015, Keim2019}. Until recently, the largest, uniform sample of these PS was produced by \cite{Callingham2017}. This sample contained sources peaking between 72--700\,MHz, and provided unprecedented spectral coverage below the spectral turnover, yet owing to the relatively high flux density limit, the sample had a low density of only 0.06 $\text{sources/deg}^2$. Table~\ref{tab:literature-samples} compares that work to other recent samples from the literature, as well as to our new sample presented here. We note that of these samples, the first to follow \citet{Callingham2017} was the \cite{Ross2021} study of low-frequency variability which has a lower source density again owing to the requirement for multiple GLEAM epochs. This was followed closely by the work of \citet{Slob2022} using the LOw Frequency ARray (LOFAR: \citealt{vanHaarlemLOFAR}) to identify a comparative sample of PS sources in the northern equatorial sky. This was recently superseded by the work of \citet{Bailieux2024}, who identified several thousand PS sources in the northern sky by combining survey data from LOFAR, and the Very Large Array (VLA) spanning 60--160\,MHz (LOFAR -- the LOFAR Two-metre Sky Survey) and 1.4/3\,GHz (VLA -- the VLA Sky Survey and NRAO VLA Sky Survey) in separate subsamples. However, a comprehensive sample of PS sources, with peaks spanning a broader frequency range and extending to higher frequencies, is crucial for any statistical study of radio AGN evolution, since the linear size of a PS source (a proxy for its age) is expected to be inversely proportional to the restframe frequency of its spectral peak \citep{Snellen2000, deVries2009, Jeyakumar2016}. 

This paper presents the first such sample of PS sources selected initially using a single, narrow frequency band, yet with spectral peaks spanning 70\,MHz -- 20\,GHz in the observers' frame. This use of a single selection frequency, along with the application of our Bayesian modelling framework \textsc{RadioSED}, and our focus on a well-defined region of sky, means that our sample completeness can be very well constrained. In fact, this sample of {\fullPScount} sources is at least 90 per cent complete down to 200\,mJy across the full range of observed peak frequencies captured. We focus our search on the 300\,deg$^2$ Sloan Digital Sky Survey (SDSS) Stripe 82 field \citep{Abazajian2009}, a region rich with multiwavelength coverage from radio frequencies through to TeV $\gamma$-rays. Despite this abundance of data, only eleven PS sources have been identified in the field to date, as shown in Figure \ref{fig:skydist-sample}. It is therefore a useful test of \textsc{RadioSED} to apply it to this well-studied field, and the resulting sources in turn form a clean sample with which to probe the properties of PS radio galaxies. Section \ref{sec:sample-definition} provides further details as to the composition of our sample, Section~\ref{sec:radio} presents the radio properties of the sample, including measures of source structure and variability, in Section~\ref{sec:host_optical} we consider the optical and IR properties of the hosts of our PS sources, and finally in Section~\ref{sec:PS-population} we return again to the radio properties of our sources in the context of galaxy evolution. Conclusions and a summary of our findings are presented in Section~\ref{sec:conclusions}.

Throughout this work, we adopt a flat, $\Lambda$ cold dark matter ($\Lambda$CDM) cosmology in line with values from \citet{Planck2016}; $\Omega_{\rm{m}}$ = 0.308, $\Omega_\Lambda = 1-0.308$, and $H_0 =67.8\text{km\,s}^{-1}\text{Mpc}^{-1}$.

\section{The sample: peaked spectrum radio sources in stripe 82}\label{sec:sample-definition}

The initial construction of our sample is intentionally simple; we make cuts in position and flux density on a parent sample of radio sources, and then perform radio spectral fitting. Using the 888\,MHz Rapid ASKAP Continuum Survey \citep[RACS-low;][]{Hale2021} catalogue as our reference, we select out only those sources within Stripe 82 ($300^{\circ} < \alpha < 60^{\circ}$, $-1.26^{\circ} < \delta < 1.26^{\circ}$), and which have an integrated flux density $S_{\text{888\,MHz}} > 10$\,mJy. This flux density cut ensures our parent sample is at least 95 per cent complete (see Section 6 of \citealt{Hale2021}), whilst simultaneously removing the many star-forming galaxies which dominate the microJansky radio sky, and which might otherwise contaminate our final AGN sample \citep[e.g.][]{Padovani2016}. This leaves us with 7\,251 radio sources in Stripe 82, all of which were run through \textsc{RadioSED} to construct and then fit their broadband radio spectral energy distributions. The distribution of peaked sources from amongst this sample is shown Figure \ref{fig:skydist-sample}, where we have overplotted the previously published PS sources as purple stars, to illustrate the expansion in sample size achieved with this work.

\begin{figure*}
    \centering
    \includegraphics[trim={1cm 0cm 7cm 0},width=0.9\linewidth]{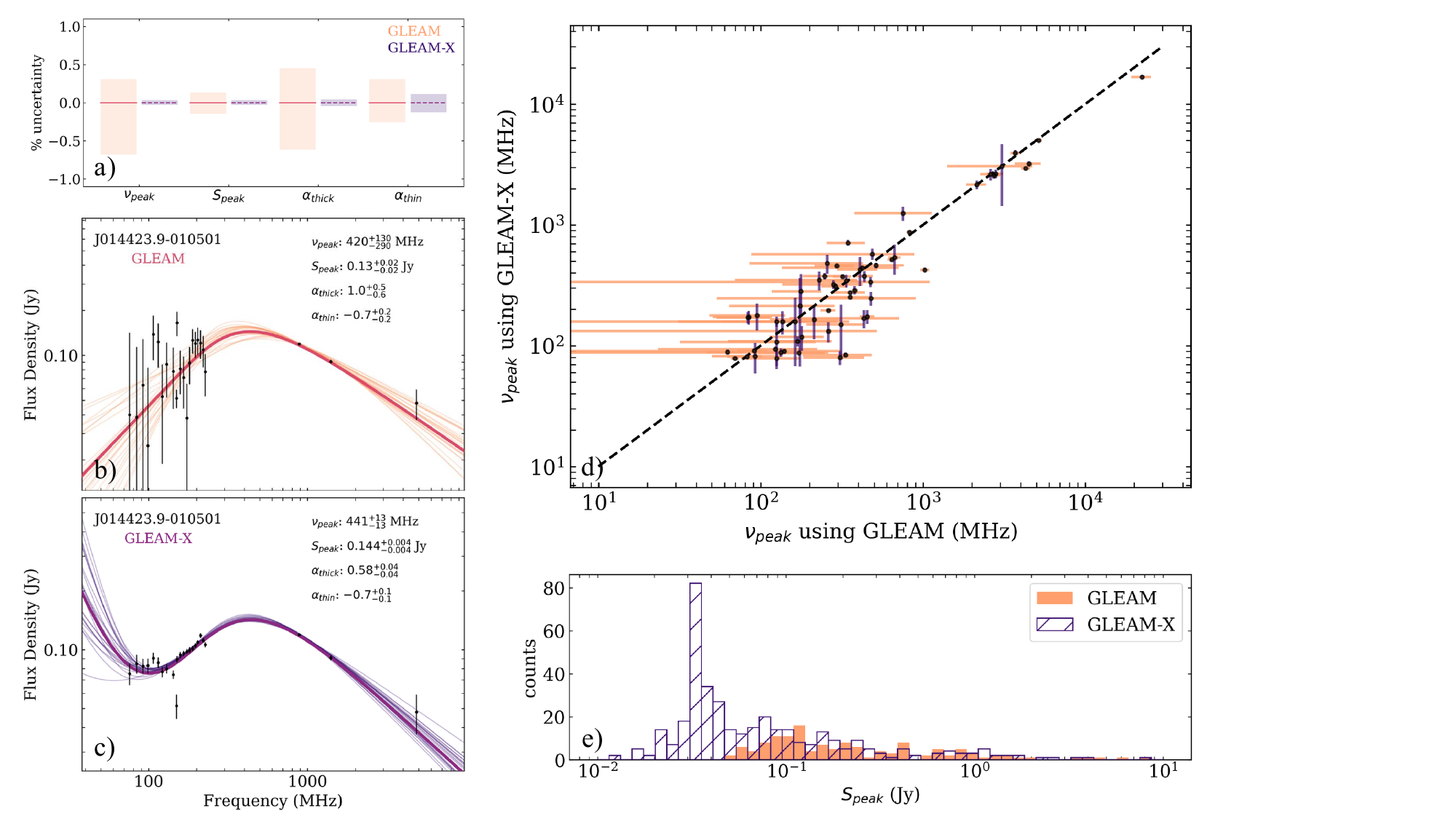}
    \caption{A summary of the accuracy of \textsc{RadioSED} fits when using GLEAM data (orange) as opposed to data from the second release of GLEAM-X (purple). Panel a) shows the magnitude of the errors on best fit parameters for a typical peaked spectrum source in our sample. The solid, orange lines are set to zero and represent the mean parameter values derived using GLEAM data, the orange shaded regions indicating the fractional uncertainty on each of these. The dotted purple lines indicate the mean parameter values derived using GLEAM-X data relative to the GLEAM-derived parameters, with purple shading indicating the (significantly smaller) fractional uncertainties on these. Panels b) and c) show the best fit model for this same source when using GLEAM data (panel b), orange) as opposed to GLEAM-X (panel c), purple). The values of the fit parameters are inset in these panels as text. Panel d) shows the peak frequency and associated uncertainty derived from the best fitting \textsc{RadioSED} model. This is for all sources in Stripe 82 classified as peaked spectrum using both GLEAM and GLEAM-X data, with the black dotted line indicating a 1-1 relationship. Panel e) shows the number of PS sources identified using each catalogue as a function of their peak flux density ($S_\text{peak}$)}.
    \label{fig:gleamx-gleam-compare}
\end{figure*}

\begin{table}
	\centering
	\caption{Parameters for several of the most recent PS samples from the literature, compared to those presented in this work.}
	\label{tab:literature-samples}
	\begin{tabular}
{m{12mm}m{6mm}m{10mm}m{10mm}m{28mm}}
		\hline
		  Selection Frequency & Sample Size & Survey Area & Source Density & Reference\\
           (MHz) & & (deg$^2$) & (deg$^{-2}$) & \\
		\hline
            \hline
            $72-230$ & 1,483 & 24,831 & 0.06 & \citet{Callingham2017} \\ 
            $72-230$ & 123 & 8,000 & 0.015 & \citet{Ross2021} \\ 
            $42-1400$ & 373 & 740 & 0.50 & \citet{Slob2022} \\ 
            $120-3000$ & 8,032 & 5,635 & 1.43 & \citet{Bailieux2024} \\ 
            $42-1400$ & 506 & 740 & 0.68 & \citet{Bailieux2024} \\ 
            888 & \fullPScount & 300 & 1.20 & This work \\
		  
            \hline
	\end{tabular}
\end{table}

\subsection{Radio spectral fitting}\label{sec:fitting}
To identify the peaked spectrum sources from amongst this parent catalogue, we first fit them using \textsc{RadioSED} with the default selection of radio surveys as outlined in \citetalias{Kerrison2024}, including the GaLactic and Extragalactic All-sky Murchison Widefield Array (GLEAM) survey at the lowest frequencies \citep{Hurley-Walker2017, Tingay2013TheFrequencies, Wayth2018}. However, since the publication of \citetalias{Kerrison2024}, the GLEAM eXtended survey (GLEAM-X) DR II has been released which covers Stripe 82 eastwards of $\alpha > 310^{\circ}$, and pushes down the 95 per cent completeness level to 16\,mJy, over an order of magnitude fainter than the original GLEAM catalogue \citep{HurleyWalkerGLEAMX,Ross2024}. Accordingly, we re-ran our SED fitting using this new, low frequency catalogue with a statistically-selected match radius of 12 arcseconds, determined using the method outlined in \citetalias{Kerrison2024}, to investigate any potential enhancements to the sample. As expected, this led to a significant improvement in both the size of the peaked spectrum sample, and in the precision of individual, peaked spectrum fits. A summary of these improvements is shown graphically in Figure \ref{fig:gleamx-gleam-compare}. 

More specifically, incorporating GLEAM-X data reduced the uncertainty in fit parameters from 34 per cent to 13 per cent on average across all peaked spectrum sources common to both fitting runs. In many sources for which the optically thin slope is solely constrained by either GLEAM or GLEAM-X data, this reduction is even more significant, as in the case of J014423.9--010501 shown in panels a)--c) of Figure \ref{fig:gleamx-gleam-compare}, where the more sensitive GLEAM-X data reveals a low-frequency flattening, possibly indicative of an older epoch of activity. The reduction in uncertainties across the broader sample is shown for model-derived peak frequencies ($\nu_\text{peak}$) in panel d), where it is shown to be consistent across the two runs to within the larger uncertainties from the GLEAM-derived models. Our ability to constrain the peak frequency is of particular importance, as this has been shown to correlate directly with the linear size, and hence, the age of a source \citep{ODea1998}. In addition to the precision of individual fits, the size of our peaked spectrum sample increased fourfold, from 92 to \nocensoredPScount\,with the use of GLEAM-X data, extending down to sources with a peak flux $11.7\pm1$\,mJy, as shown in panel e). This is an exciting hint of what is to come with the larger, peaked spectrum sample of Ross et al. (in preparation) covering the full, GLEAM-X DRII region. And it is for these reasons that we proceed with the sample derived using the GLEAM-X data for the remainder of this paper.

\begin{figure*}
    \centering
    \includegraphics[width=1.0\linewidth, trim={0.1cm 10.5cm 0.1cm 0},clip]{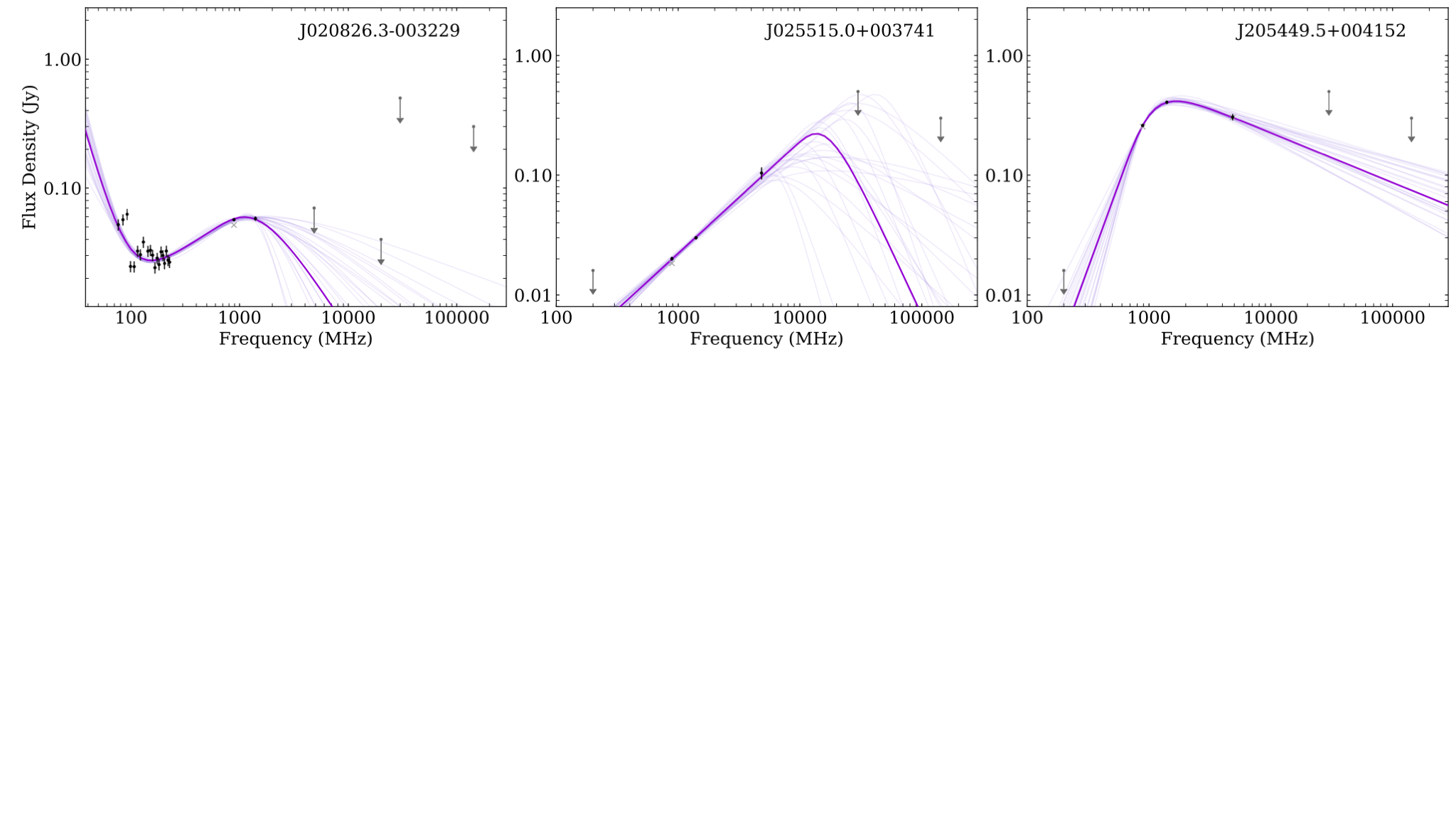}
    \caption{Examples of sources for which the peak can be identified only with the addition of censored datapoints from the surveys outlined in Table \ref{tab:survey_limits}. Upper limits are indicated by arrows. The best fit model is shown as a purple, solid line, with lighter lines indicating draws from the posterior.}
    \label{fig:censored_seds}
\end{figure*}

Since we are focused on a well-defined region of sky, we were also able to incorporate survey limits into our fitting as censored data points within individual SEDs. These additional constraints are outlined in Table~\ref{tab:survey_limits}, where the limiting flux densities are taken as the 95 per cent completeness limits drawn from the respective papers. The implementation of censored fitting itself was described in \citetalias{Kerrison2024}. These upper limits led to the inclusion of thirteen additional PS sources in our final sample, and improved constraints on the spectral properties of several more sources in which a peak had already been identified. Examples of sources which benefit from the use of censored data are shown in Figure~\ref{fig:censored_seds}. Overall, our sample therefore comprises \fullPScount\,PS sources within the Stripe 82 region. The fit parameters of these PS sources are given in Appendix~\ref{sec:appxA-sedparams}, and the full catalogue is available in the online version of this paper. 

For a source to be classified as PS, it must have observations either side of the predicted peak, and it must have fractional error $\frac{\Delta\alpha_{\text{thick}}}{\alpha_{\text{thick}}} < 1$ to avoid ambiguity in spectral shape (where $\alpha_{\text{thick}}$ is the spectral index below the broadband peak). Furthermore, as discussed in \citetalias{Kerrison2024}, we have separated out those sources which have spectral indices shallower than required for the canonical definition of a PS source (i.e. $ -0.5 < \alpha_{\text{thin}} < 0$ or $ 0 < \alpha_{\text{thick}} < 0.5 $) into a ``soft peaked spectrum'' (SPS) category within the catalogue. This is for purely historical reasons, to allow for easy comparison with older PS samples from the literature. For our analysis here though, the PS and SPS sources will be collectively analysed as, and assumed to be, peaked spectrum sources. 

Overall, from the 7\,251 radio sources within our selected field, 4\,458 could not be fit due to an insufficient number of flux density measurements or meaningful constraints provided by upper limits. This population of unfitted sources has a median flux density of 18.04\,mJy, and 90 per cent have a flux below 70\,mJy, so they comprise mainly faint sources in the field. 

From amongst the sources which could be reliably fit, PS sources make up approximately 13 per cent, which falls comfortably within the range of population fractions calculated by \cite{Bailieux2024} at one end (3.9 per cent for sources peaking at MHz frequencies) and \cite{ODea1998} at the other (20 per cent typically peaking at GHz frequencies). The SED plots for the full sample of \fullPScount\,PS sources can be found in the supplementary material online.

\begin{table}
	\centering
	\caption{The survey limits incorporated into \textsc{RadioSED} for the identification of the PS sample in this paper, in addition to those already described in \citetalias{Kerrison2024}. The chosen flux density is the 95 per cent completeness limit of the relevant survey, and if it applies only to a sub-section of Stripe 82, this constraint is given in the table.}
	\label{tab:survey_limits}
	\begin{tabular}
{m{13mm}m{9mm}m{7mm}m{12mm}m{24mm}}
		\hline
		  Survey & Frequency & Limit & Region of& Reference\\
           & (GHz) & (Jy) & Stripe 82 & \\
		\hline
            \hline
            GLEAM-X & 0.2 & 0.016 & $\alpha > 310^{\circ}$ & \cite{Ross2024} \\
            PMN & 4.85 & 0.07 & full & \cite{Griffith1993TheReduction}\\
            AT20G & 20 & 0.04 & $\alpha < 200^{\circ}$, $\delta < 0^{\circ}$ & \cite{Massardi2011}\\
            PCCS & 30 & 0.5 & full & \cite{Planck2013}\\
            PCCS & 143 & 0.3 & full & \cite{Planck2013}\\
		  
            \hline
	\end{tabular}
\end{table}

\subsection{Completeness: sample limits}\label{sec:completeness-synthetic}

\begin{figure}
    \centering
    \includegraphics[width=1.0\linewidth]{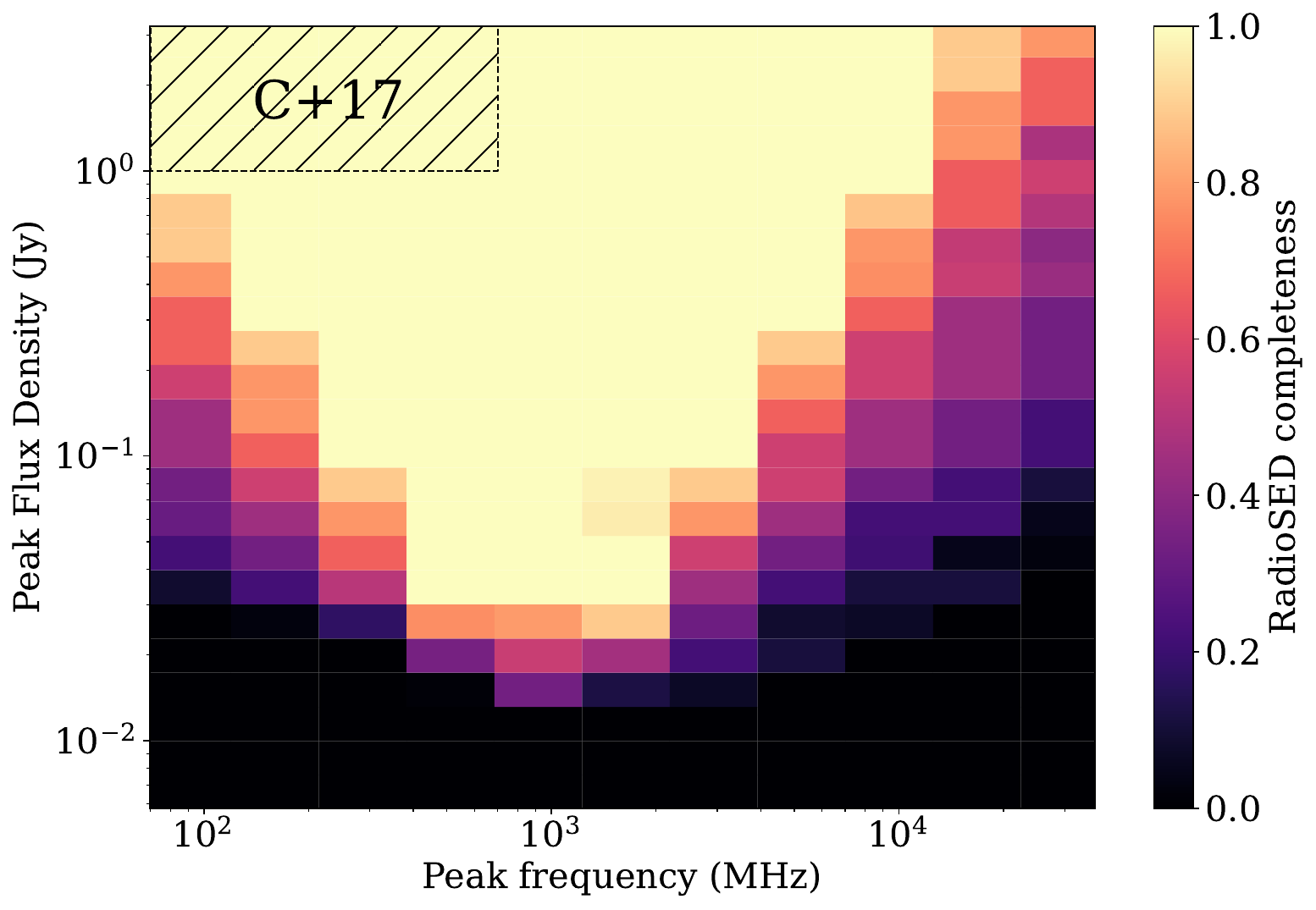}
    \caption{The completeness of our PS sample as a function of peak flux density and peak frequency, divided into logarithmically-spaced bins. Completeness calculations were performed by feeding synthetic spectra to \textsc{RadioSED}, and using the set of parameters specified in Section~\ref{sec:fitting}. The hatched region indicates the area of parameter space within which the all-sky sample of \citet{Callingham2017} was estimated to be complete.}
    \label{fig:completeness_sims}
\end{figure}

Although our initial selection criterion was to impose a single flux density cut (10\,mJy at 888\,MHz), the completeness of our sample will be a function of both peak frequency and peak flux density, since a source must be detected in at least three unique frequency bands to be fit. Within the region of GLEAM-X DR II (eastwards of $\alpha = 310^{\circ}$), we estimate the completeness of our PS sample using a simple Monte Carlo test. We generate just over 23\,000 synthetic spectra with spectral indices uniformly sampling the interval $ 0.5 < |\alpha| < 2.1$, peak frequencies logarithmically sampling the interval between 50\,MHz --30\,GHz, and peak flux densities spanning $5$\,mJy -- $2.8$\,Jy, also logarithmically spaced. For each combination of spectral parameters, we generate flux density measurements at frequencies corresponding to the surveys used in fitting, and keep only those above the 95 per cent completeness limit in each survey, otherwise replacing them with upper limits where these were used in constructing our sample (we refer the reader to \citetalias{Kerrison2024}, Table 1 for further information on these surveys). The resulting synthetic spectra are fit with \textsc{RadioSED}, and the criteria from Section~\ref{sec:fitting} are applied to determine which would fall into the `PS' category as defined here. The results of this test are summarised in Figure~\ref{fig:completeness_sims}, where the colour gradient indicates the expected completeness from these simulations.

Evidently, our sample completeness varies as a function of both peak frequency and peak flux density, and even at our selection frequency of 888\,MHz we do not reach 100 per cent completeness at 10\,mJy. This is undoubtedly due to the higher flux density limits of the other surveys required for fitting, which cannot adequately sample spectra which peak at or below a few tens of mJy. Nevertheless, we are most complete to sources peaking at or around our 888\,MHz selection frequency. These simulations also suggest that our sample is complete across a large fraction of the peak parameter space. For comparison, we overplot the hatched region in Figure~\ref{fig:completeness_sims} to indicate the completeness of the \cite{Callingham2017} sample. To our knowledge, this is the only other PS sample to date for which completeness statistics have been calculated, and it is clear that the present sample promises a vast improvement in the diversity of PS sources we can detect. The inclusion of additional, high-frequency surveys, along with the more sensitive GLEAM-X data, and censored fitting with upper limits all allow us to identify PS sources that are fainter and peak at higher frequencies than the \cite{Callingham2017} sample. The former of these improvements may lead to the identification of PS sources that are either intrinsically less powerful or more distant, while the latter suggests we are sensitive to a population in which the jets are even more compact. This sample may be therefore thought of as a narrow, deep companion to the wide-area and shallow sample in that work, and a precursor to the large-area samples that will be possible with future GLEAM-X releases, and eventually, with data from SKA--low and mid.

\subsection{Completeness: recovering published sources}\label{sec:published_ps}

Another measure of the completeness of our sample is our ability to recover known peaked spectrum sources. Guided by the references in \citet{Callingham2017} and \citet{Bailieux2024}, we searched a number of papers for peaked spectrum sources within the Stripe 82 field boundaries. These include: \citet{Stanghellini1997, ODea1998,Dallacasa2000, EdwardsTingay2004, Tinti2005HighSources, Torniainen2007, Labiano2007, Randall2011,  Callingham2017} and \citet{Slob2022}. After removing duplicates and those sources classified as ``Compact Steep Spectrum'' (i.e. without any observed spectral turnover), we were left with the eleven sources outlined in Table~\ref{tab:s82_known_ps}.

\begin{figure*}
    \includegraphics[width=\textwidth, trim={1.1cm 1.5cm 0.7cm 0},clip]{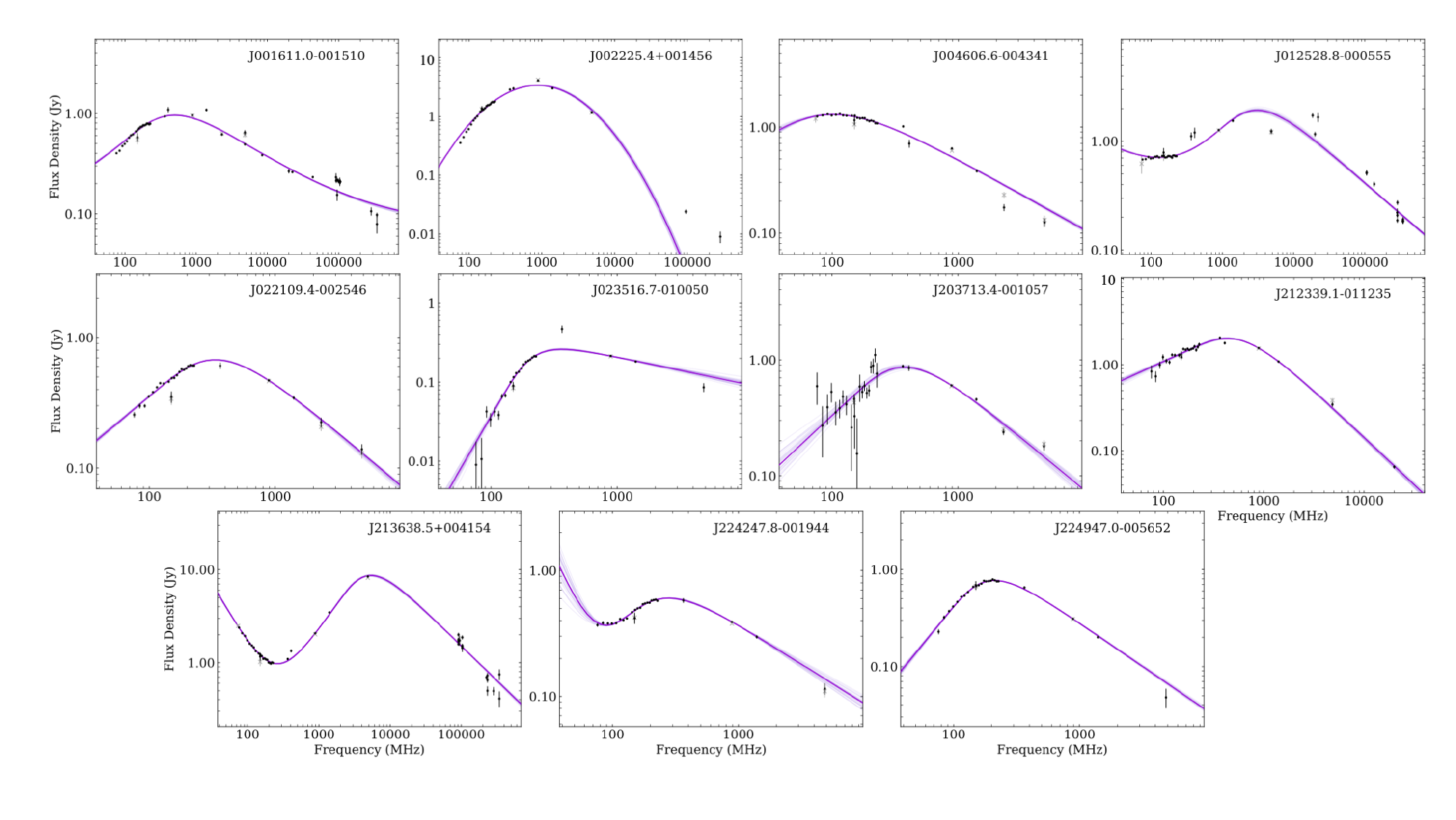}
    \caption{SEDs of all eleven known peaked spectrum sources from the Stripe 82 field obtained using \textsc{RadioSED}. The dark line in each plot shows the best fit of the most probable model from \textsc{RadioSED}, while the array of lighter lines represents 25 draws from the posterior of this model.}
    \label{fig:s82_ps}
\end{figure*}

We recovered peaked SEDs for all eleven previously known PS sources within our sample. The fit parameters output by \textsc{RadioSED} are given in Table~\ref{tab:s82_known_ps}, to aid comparing our results to those from the literature. The SEDs for these sources are shown in Figure~\ref{fig:s82_ps}. 

Of these eleven sources, \text{J023516.7--010050} would canonically be classified as flat spectrum since it has $\alpha_\text{thin} > -0.5$, while \text{J001611.0--001510} is on the borderline of a flat spectrum classification. However, as can be seen from Figure~\ref{fig:s82_ps} these are well fit by PS models, and our spectral indices agree broadly with those derived by \cite{Callingham2017}. Accordingly \text{J023516.7--010050} is classified as ``soft peaked spectrum'' for compatibility with previous definitions. The other nine sources show clear peaked spectrum SEDs, although \text{J001611.0--001510}, \text{J012528.8--000555}, and \text{213638.5+004154} appear in the 5th edition of the Roma-BZCAT blazar catalogue as ``Flat Spectrum Radio Quasars'' \citep{massaro2015}, calling into question whether they are true Peaked Spectrum sources. We consider this question further in Section \ref{sec:blazars}, but ultimately we keep these sources in our PS sample, as they do not possess a flat spectrum, a key component of their classification in Roma-BZCAT.

While the spectral parameters derived by \textsc{RadioSED} are not always in close agreement with those from the literature, they appear to appropriately characterise the SED shapes in Figure~\ref{fig:s82_ps}. Some discrepancy between parameters from the literature and \textsc{RadioSED} is evidently due to the broader frequency coverage of observations used in this work compared to previous approaches, but a deeper analysis beyond this is challenging, as many of the fits from older works use data that was not published in tabular form. However, we demonstrated in \citetalias{Kerrison2024} that \textsc{RadioSED} can accurately recover injected parameters in synthetic fits, so any discrepancy is real, and must be rooted in the subtleties either of the data used, the fitting methods, or both. 

\begin{landscape}
\bgroup
\defcitealias{Callingham2017}{C17}
\defcitealias{Torniainen2007}{T07}
\defcitealias{ODea1998}{O98}
\defcitealias{EdwardsTingay2004}{E04}
\defcitealias{Spoelstra1985}{S85}
\begin{table}
    \centering
	\caption{Peaked Spectrum sources within the Stripe 82 field that were previously identified in the literature.* Source names are from the RACS-low catalogue, and literature values for model parameters (peak frequency: $\nu_\text{p, lit}$, peak flux: $S_\text{p, lit}$, and optically thick and thin spectral indices: $\alpha_\text{thick, lit}$, $\alpha_\text{thin, lit}$) are drawn from the citations listed, with uncertainties quoted where available. All other model parameters are derived from \textsc{RadioSED}. Citations are not exhaustive for each source, but represent those works in which the source was first identified as peaked spectrum, and from which the literature parameters were derived. The citations in Column 11 are as follows: S85: \citet{Spoelstra1985}, O98: \citet{ODea1998}, E04: \citet{EdwardsTingay2004}, T07: \citet{Torniainen2007}, and C17: \citet{Callingham2017}. Where a model parameter is unknown or not applicable to a particular source, that column is marked with a dash.}
	\label{tab:s82_known_ps}
	\begin{tabular}{lccccllllcr} 
		\hline
		  Name  & $\nu_\text{p, lit}$ & $S_\text{p, lit}$& $\alpha_\text{thick, lit}$ & $\alpha_\text{thin, lit}$ & $\nu_\text{p}$ & $S_\text{p}$ & $\alpha_\text{thick}$ & $\alpha_\text{thin}$ & $\alpha_\text{retrig}$& Refs.\\
            & (GHz) & (Jy) & & & (GHz) & (Jy) & \\
		\hline
            \hline
J001611.0--001510 & >0.843 & -- & $0.42\pm0.08$ & -- & $0.526^{+0.015}_{-0.015}$ & $0.885^{+0.005}_{-0.005}$ & $0.67^{+0.01}_{-0.01}$ & $-0.52^{+0.01}_{-0.01}$ & $-0.002^{+0.006}_{-0.006}$ & \citetalias{Callingham2017} \\
J002225.4+001456 & 0.76 & -- & 0.46 & -- & $0.873^{+0.012}_{-0.012}$ & $3.39^{+0.03}_{-0.03}$ & $1.59^{+0.006}_{-0.007}$ & $-3.77^{+0.05}_{-0.05}$ & -- & \citetalias{ODea1998, Torniainen2007} \\
J004606.6--004341 & $0.09\pm0.04$ & $1.41\pm0.19$ & -- & $-0.59\pm0.09$ & $0.117^{+0.006}_{-0.005}$ & $1.312^{+0.006}_{-0.009}$ & $0.69^{+0.07}_{-0.06}$ & $-0.66^{+0.01}_{-0.01}$ & -- & \citetalias{Callingham2017} \\
J012528.8--000555 & >0.843 & -- & $0.46\pm0.09$ & -- & $3.31^{+0.17}_{-0.17}$ & $1.79^{+0.04}_{-0.05}$ & $0.77^{+0.03}_{-0.03}$ & $-0.59^{+0.01}_{-0.01}$ & $-0.38^{+0.03}_{-0.03}$ & \citetalias{Callingham2017} \\
J022109.4--002546 & $0.35\pm0.07$ & $0.72\pm0.08$ & $0.90\pm0.21$ & $-0.91\pm0.32$ & $0.385^{+0.012}_{-0.011}$ & $0.667^{+0.003}_{-0.003}$ & $0.81^{+0.02}_{-0.02}$ & $-0.82^{+0.02}_{-0.02}$ & -- & \citetalias{Callingham2017} \\
J023516.7--010050 & $0.31\pm0.08$ & $0.27\pm0.11$ & $3.62\pm1.85$ & $-0.34\pm0.34$ & $0.223^{+0.007}_{-0.006}$ & $0.215^{+0.007}_{-0.007}$ & $2.76^{+0.13}_{-0.11}$ & $-0.33^{+0.03}_{-0.03}$ & -- & \citetalias{Callingham2017} \\
J203713.4--001057 & $0.40\pm0.08$ & $0.86\pm0.05$ & $0.83\pm0.26$ & $-0.96\pm0.24$ & $0.404^{+0.038}_{-0.033}$ & $0.863^{+0.025}_{-0.027}$ & $1.02^{+0.14}_{-0.13}$ & $-0.89^{+0.04}_{-0.04}$ & -- & \citetalias{Callingham2017} \\
J212339.1--011235 & $0.34\pm0.07$ & $1.75\pm0.05$ & $0.71\pm0.14$ & $-0.67\pm0.14$ & $0.648^{+0.026}_{-0.025}$ & $1.85^{+0.04}_{-0.04}$ & $0.53^{+0.03}_{-0.03}$ & $-1.1^{+0.01}_{-0.02}$ & -- & \citetalias{Spoelstra1985, Torniainen2007, Callingham2017} \\
J213638.5+004154 & 5.9 & -- & 1.6 & $-0.88\pm0.08$ & $5.28^{+0.18}_{-0.18}$ & $8.65^{+0.11}_{-0.1}$ & $1.08^{+0.02}_{-0.02}$ & $-0.75^{+0.01}_{-0.01}$ & $-1.31^{+0.02}_{-0.02}$ & \citetalias{ODea1998, EdwardsTingay2004, Callingham2017} \\
J224247.8--001944 & $0.44\pm0.15$ & $0.77\pm0.18$ & $0.52\pm0.18$ & $-1.56\pm0.90$ & $0.242^{+0.014}_{-0.012}$ & $0.583^{+0.012}_{-0.015}$ & $1.4^{+0.14}_{-0.13}$ & $-0.63^{+0.03}_{-0.04}$ & $-2.3^{+0.4}_{-0.4}$ & \citetalias{Callingham2017} \\
J224947.0--005652 & $0.26\pm0.09$ & $0.79\pm0.05$ & $0.92\pm0.29$ & $-1.18\pm0.25$ & $0.196^{+0.003}_{-0.003}$ & $0.77^{+0.002}_{-0.002}$ & $1.59^{+0.04}_{-0.04}$ & $-0.89^{+0.01}_{-0.01}$ & -- & \citetalias{Callingham2017} \\
		\hline
	\end{tabular}
\end{table}
\begin{tablenotes}
\item[*] *During the preparation of this manuscript, \citet{Sun2025} was published presenting a new analysis of PS sources, primarily using GLEAM data. These PS sources are not included in this analysis.
\end{tablenotes}
\egroup
\end{landscape}

\subsection{Reliability}\label{sec:reliability}

The reliability of a new sample of PS radio AGNs is difficult to determine a priori. In Section 4.2 of \citetalias{Kerrison2024}, we reasoned that we might expect a misclassification rate of approximately 4 per cent in any sample of PS sources found using \textsc{RadioSED}, based on simulations. However, this assumed a certain degree of variability amongst flat spectrum radio sources, and that they comprise a certain fraction of the radio population --- neither of which is particularly well constrained in the literature. We might therefore take this 4 per cent as a fiducial contamination rate in our sample, which would suggest some 14 of our sources are not true PS AGNs. However, to say more requires a deeper, multiwavelength consideration of the sources in question, which we will return to again in the context of variability in Section~\ref{sec:variability}.

\subsection{Blazar contamination}\label{sec:blazars}

At the highest level, a blazar is an AGN in which the jet axis is aligned along our line of sight such that its emissions are relativistically beamed. This produces apparent superluminal motion, and causes the source flux density to vary rapidly across the full electromagnetic spectrum. In a gigahertz radio image with arcsecond resolution, a blazar will therefore appear as a compact source. Although a PS radio AGN could of course have its jets aligned in this way, the likelihood of observing such an object is probably very low, and in any case its emissions would be dominated by the blazar-like characteristics of its aligned jet. For this reason, those working on PS sources tend to consider blazars a separate class of object \citep[e.g.][]{Orienti2007, ODea2021}. By contrast, the original Roma-BZCAT catalogue of blazars included PS sources peaking at or above 1\,GHz \citep[][Sec. 3]{Massaro2009}, and these have propagated through into the latest version of that work. More recently, \citet{Behiri2025} showed that a number of Fermi blazars in fact have broadband radio spectra with complex shapes including peaked and low-frequency upturns indicative of episodic activity. Furthermore, many works identifying PS sources (including this one) have necessarily had to combine observations from different epochs in order to obtain the broadband spectral coverage necessary for classification. The dynamic variability of blazars combined with this heterogeneous data makes any such sample somewhat susceptible to blazar contamination, though such contamination should be naturally reduced in the coming years with the publication of more contemporaneous, widefield, and broadband surveys from the SKA and its precursors and pathfinders. Nevertheless, at present this makes separating out blazars from candidate PS radio AGNs a difficult task using blazar catalogues alone.

We have already seen in Section~\ref{sec:published_ps} that three of the eleven previously-published PS sources in our sample are found within the Roma-BZCAT catalogue of blazars. However, in each case they are classed as ``Flat Spectrum Radio Quasars'' (FSRQs), a classification which is typically based on only two flux density measurements, plus an optical spectrum indicative of a quasar. A crossmatch between the 5th edition of the Roma-BZCAT and the PS sources in Stripe 82 reveals a further 13 matches, but again 10 of these are classified as FSRQs, with only the final three exhibiting featureless optical spectra of a BL Lac type, clearly indicative of a blazar \citep{massaro2015}. These three are excluded from further analysis as PS sources. Since it is known that PS sources can be hosted by both galaxies and quasars, we do not discount the 10 FSRQ sources from our sample on this basis alone, though we do note that a visual inspection of the SEDs of \text{J001611.0--001510}, \text{J012528.8--000555}, and \text{213638.5+004154} in Figure~\ref{fig:s82_ps} (the previously-published PS sources also in the Roma-BZCAT) reveals some scatter around the best fit model in each case, which may be indicative of radio variability consistent with episodic blazar activity. 

Finally, since the Roma-BZCAT relies on both optical spectra and radio flux density for classification, it will not include those sources in our sample lacking optical spectra. This may mean there are additional blazar candidates contaminating our sample which cannot be identified by this method alone. We leave further consideration of potential blazar contamination to Section~\ref{sec:variability}, where we consider the other key metric by which blazars are typically identified; variability.

\section{Radio properties}\label{sec:radio}

PS sources are expected to be compact, with linear sizes $\lesssim 1$\,kpc, corresponding to angular scales $\theta \sim 1.2$\,arcsec at $z=1$, and to have a radio spectrum with a low degree of variability. Here, we study the sub-arcsecond compactness of our sample using relatively novel compactness measures at 162\,MHz and 20\,GHz. We also consider the variability of our sample at $\nu \sim 800\,$MHz and consider the possibility of both intrinsic variation, and flux modulation by an external scattering screen. The radio frequency data used in the following subsections can be found in Appendix~\ref{sec:appxB-radioparams}, with the full version available online.

\subsection{Source Structure from continuum surveys}\label{sec:large-structure}

To obtain first-order constraints on the source structure we measure compactness based on two of the surveys used in SED fitting. Following the procedure used in survey description papers such as \citet{Bondi2008, Shimwell2019}, and \citet{Hale2021}, we determine which of our sources are unresolved up to the resolution limit of both RACS-low (15\,arcsec at 888\,MHz) and GLEAM-X ($\sim45$\,arcsec at 200\,MHz), two of the surveys with most complete coverage for our sample. We construct an envelope within which unresolved sources are expected to reside when the ratio of their total to peak fluxes ($F_{\rm{tot}}/F_{\rm{pk}}$) is plotted against their signal to noise (SNR). In the case of RACS-low, this envelope was already derived in \citet{Hale2021} Section 5.2.1, and we use those same parameters here. In the case of GLEAM-X DRII, we follow the method from that paper to derive an envelope of the form:
\begin{equation}
    \dfrac{F_{\rm{tot}}}{F_{\rm{pk}}}_{\pm} = 0.995 \pm 0.281\times \text{SNR}^{-0.313}
\end{equation}

\begin{figure}
\centering\includegraphics[width=0.5\textwidth]{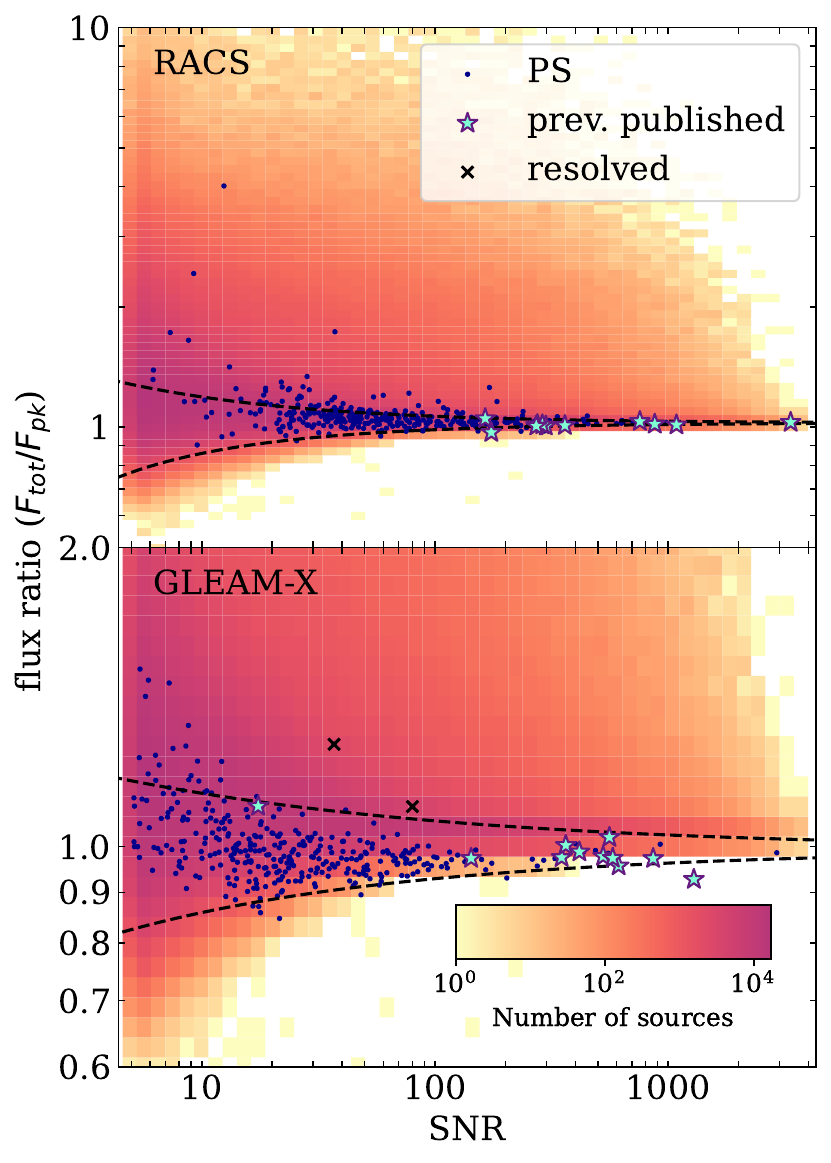}
    \caption{Heatmap of the flux ratio $\frac{F_{\rm{tot}}}{F_{\rm{pk}}}$ for all sources in RACS-low (top) and GLEAM-X (bottom), overplotted with our PS sources (navy dots) and previously-published PS sources in Stripe 82 (cyan stars). The dotted lines inscribe the envelope of unresolved sources as defined in \citet{Hale2021} Section 5.2.1.}
    \label{fig:survey-compactness-scatter}
\end{figure}

These envelopes are shown in the two panels of Figure~\ref{fig:survey-compactness-scatter}. Although it may seem at first glance that several of our sample fall outside of this envelope, its shape only takes into account the global scatter about an expected ratio, $\frac{F_{\rm{tot}}}{F_{\rm{pk}}}$ for an unresolved source, and not the uncertainty on the individual flux density measurements. In reality, the uncertainty on both $F_{\rm{tot}}$ and $F_{\rm{pk}}$ in our sources means that, to first order, all could reasonably be considered unresolved in RACS-low, and all bar two are unresolved in GLEAM-X. The two which may show extended structure at MHz frequencies ( J020137.2+010057 and J023036.5--005122) are indicated in Figure~\ref{fig:survey-compactness-scatter} by large black crosses, though a visual inspection of their GLEAM-X images does not reveal any obvious structure.

However, if we consider the length scales which are probed at the resolution of these surveys assuming a flat, $\Lambda$CDM cosmology, we are only sensitive to structures $D \geq 30$\,kpc at a redshift $z = 0.1$ (with RACS-low), increasing to structures as large as $D \geq 40$\,kpc at $z = 5.0$ (with GLEAM-X). This far exceeds the typical size of a PS source, which is thought to be between a few hundred parsecs up to $1$-$2\,$kpc, where they begin to transition to compact steep spectrum sources, in which the peak falls below the range of frequencies currently observable \citep{ODea2021}. Thus these surveys alone are not enough to properly constrain the structure of our PS sample. 

\subsection{Source structure: sub-kiloparsec scales}\label{sec:structure-parsec}

To perform further analysis on continuum source structure, Very Long Baseline Interferometry (VLBI) is often used to image milliarcsecond structure thanks to both the high frequency of observations ($\geq5$\,GHz) and the intercontinental baselines involved ($\geq10^3$\,km). However in this case, VLBI will resolve out much of the structure we are interested in, as even a 50\,mas resolution image probes length scales $0.01$--$0.3$\,kpc between redshifts $z = 0.1 $--$ 5$. Furthermore, VLBI catalogues are highly incomplete due to the stringent selection criteria typically applied to obtain telescope time, complicating sample-level inference. Ideally, we would like to be able to probe structures between these two spatial regimes of continuum surveys and VLBI, and fortunately there are still tools that can achieve this for us.

One such tool is interplanetary scintillation (IPS), a phenomenon where background radio sources with sub-arcsecond compact components (like Quasars or PS sources) exhibit variations in flux density over seconds-long timescales due to changes in the intervening solar wind \citep{Clarke1964IPS}. \citet{Jeyakumar2000Small-scaleSources} established IPS as a useful analysis technique for PS sources, using measurements at a range of solar elongations and position angles relative to the solar wind vector to measure source structure in conjunction with VLBI on different angular scales. A novel technique for widefield IPS detections was developed by \citet{Morgan2018IPS-method} for use with the MWA, the same instrument used for the GLEAM-X survey. Although this widefield survey technique does not provide measurements over a range of solar elongations and position angles like the targeted observations in \citet{Jeyakumar2000Small-scaleSources}, it does offer complete coverage of our sample above the sensitivity limit of the survey. Indeed, using data obtained at 162\,MHz, the MWA-IPS technique is sensitive to compact components with a $\sim0.3$\,arcsec angular diameter, and the strength of their scintillation is expressed as a `Normalised-Scintillation Index' \citep[`NSI';][]{Chhetri2018InterplanetaryFrequencies}, where a value of 1 indicates all of the flux is contained in the $\leq0.3$\,arcsec compact core, and values less than this indicate the fraction of the total MWA flux scintillating. \citet{Chhetri2018InterplanetaryFrequencies} used this technique to identify 632 sources exhibiting some level of IPS in a 5-minute observation spanning $900\,\text{deg}^2$ of sky. They note that a large fraction of their moderate-to-strongly scintillating (NSI $> 0.7$) sources are PS, and that all 21 sources from the \citet{Callingham2017} sample in their field have an NSI $> 0.7$. As follow-up to that work, \citet{Jaiswal2022VLBIArray} combined IPS analysis with VLBI measurements in a manner similar to the earlier \citet{Jeyakumar2000Small-scaleSources} study, while \citet{Sadler2019} identified the hosts of these scintillating sources as distant radio galaxies and quasars, with a median redshift $z \sim 1.5$, and approximately 30 per cent residing at redshift $z > 2$. 

How does this compare to our sample? Of the \fullPScount\,\,sources we have identified, 74 are included in the latest MWA-IPS catalogue (i.e. they have a $5\sigma$ detection in the continuum image \citep[][and catalogue in prep.]{Morgan2022}. Of these, 40 have secure IPS detections (at least one $5\sigma$ detection in the variability image), a further 24 have a marginal IPS detection ($<5\sigma$ in the variability image), and the final eleven have upper limits on their NSI. A $\sim20$ per cent detection rate (in the sense that a ``detection'' appears in the IPS catalogue - whether as measurement or upper limit) is not surprising given our original selection frequency for this sample was much higher than the 162\,MHz MWA observations. Indeed, of the 289 sources not detected with IPS, 286 have flux densities $S_{200\text{\,MHz}} \leq 100$\,mJy, and two of the three remaining sources (J014423.9--010501 and J204315.0--004037) have ``re-triggered'' type SEDs with a low frequency upturn indicative of older, extended emission. It is therefore unsurprising that their flux in the MWA band is resolved out at the angular resolution required for IPS.  Nevertheless, it is useful to consider how this population compares to the one studied in \citet{Chhetri2018InterplanetaryFrequencies}, which was selected purely based on IPS statistics.

\begin{figure}
    \centering
    \includegraphics[width=0.5\textwidth]{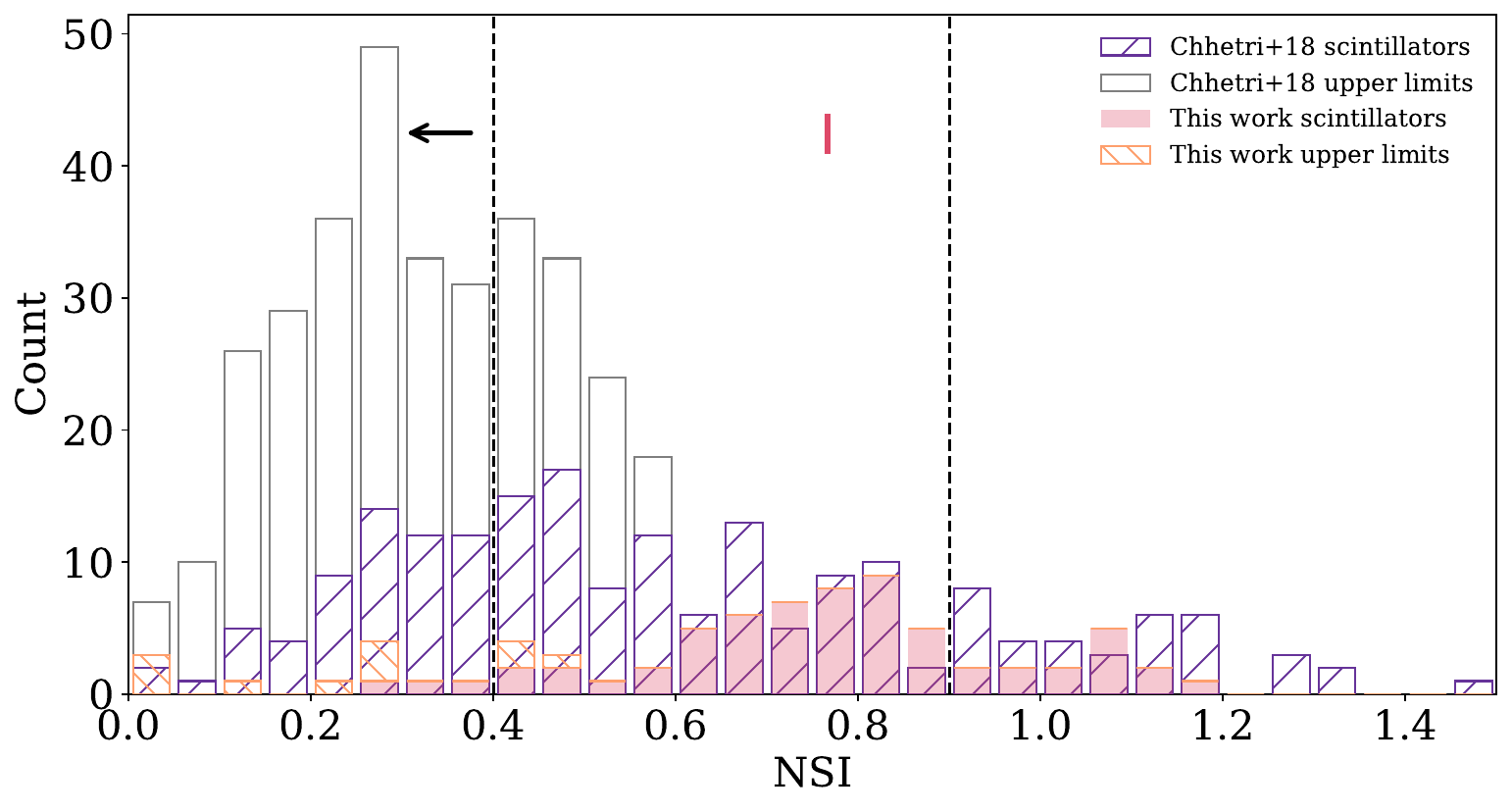}
    \caption{Distribution of NSIs for our sample (red bars) compared to the sample of \citet{Chhetri2018InterplanetaryFrequencies} (right-hatched purple bars) including upper limits (orange-hatched for our sample, black outline for the \citet{Chhetri2018InterplanetaryFrequencies} sample). The vertical dashed lines indicate the thresholds in \citet{Sadler2019} for `moderately scintillating` ($0.4 < \text{NSI} < 0.9$ and `strongly scintillating' (NSI $> 0.9$) sources. The red bar indicates the median NSI of our sample, while the black arrow indicates an upper limit on the median of the \citet{Chhetri2018InterplanetaryFrequencies} sample for reference.}
    \label{fig:compare-chhetri}
\end{figure}

In Figure~\ref{fig:compare-chhetri}, we show the distribution of NSI for our sample of PS sources, compared to the sample of \citet{Chhetri2018InterplanetaryFrequencies}, including upper limits on NSI in both cases. In our sample the distribution of NSIs is skewed towards 1 with a median of 0.77, while the \citet{Chhetri2018InterplanetaryFrequencies} reference sample contains many more censored datapoints, so that we can only estimate an upper limit of 0.38 on that sample median (we note values above 1 here are due to scatter). This is encouraging, as it suggests our sample tends to have more flux contained in sub-arcsecond structure than the general population of megahertz radio sources. Indeed, almost all of our sources are at least `moderately scintillating' by the \citet{Sadler2019} criterion (NSI $> 0.4$), indicating at least 40 per cent of their flux at 162\,MHz is contained in a compact component. However, where all 21 PS sources contained in the \citet{Chhetri2018InterplanetaryFrequencies} sample had NSI $>0.7$, only two thirds of our sample meet this threshold. This is in part due to our sample containing sources peaking at gigahertz frequencies and higher, which are typically only detected in the 162\,MHz IPS measurements if they have a low-frequency upturn indicative of older, extended emission. These findings are summarised in Table~\ref{tab:nsi_summary}.

\begin{table}
	\centering
	\caption{The distribution of NSIs for our sample, and the \citet{Chhetri2018InterplanetaryFrequencies} sample for comparison.}
	\label{tab:nsi_summary}
	\begin{tabular}{cccc}
		\hline
		  IPS & NSI & \multicolumn{2}{c}{Number (per cent)} \\
              & & This work & \citet{Chhetri2018InterplanetaryFrequencies}\\
		\hline
            Low & < 0.4 & 9 (12\%) & 59 (31\%) \\
            Moderate & 0.4 - 0.7 & 21 (28\%) & 70 (36\%)\\
            High &  > 0.7 & 45 (60\%) & 64 (33\%) \\		  
            \hline
	\end{tabular}
\end{table}

\begin{figure}
    \centering
    \includegraphics[width=0.5\textwidth]{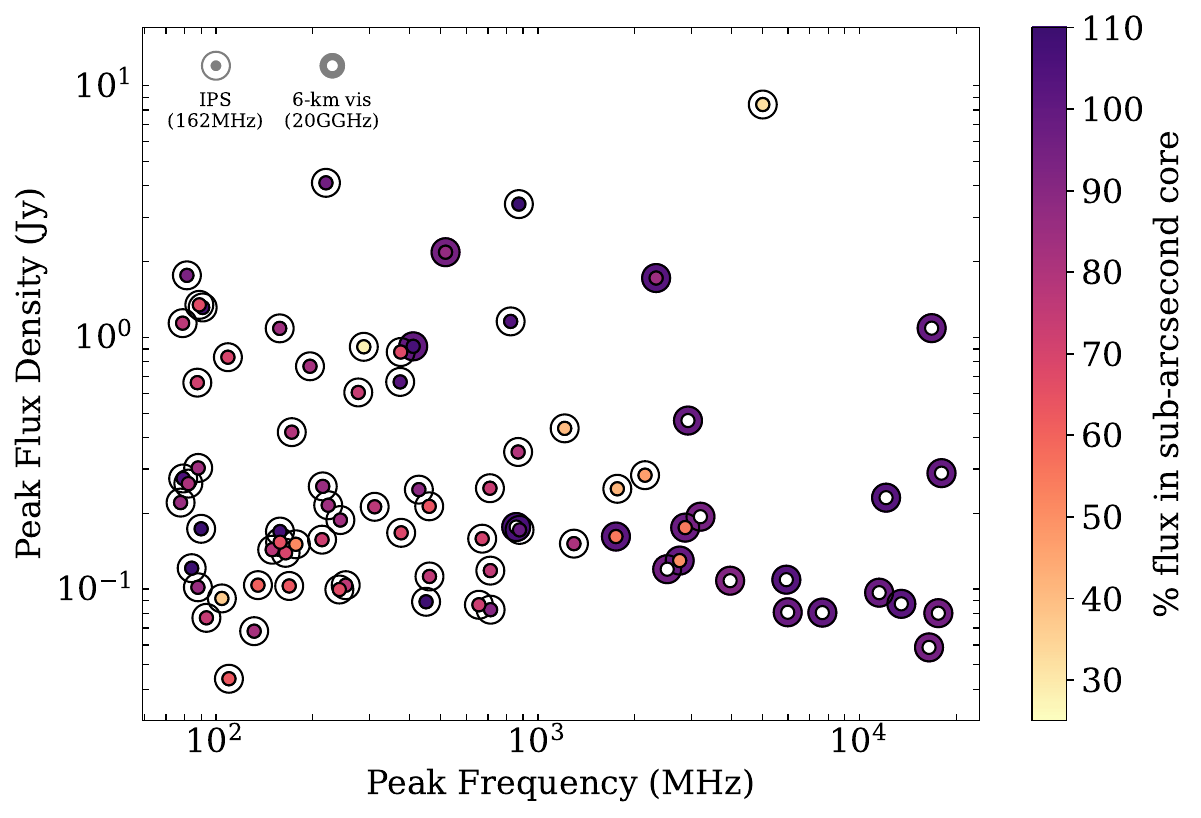}
    \caption{Sources in our full PS sample which have a sub-arcsecond compactness measure from either IPS at 162\,MHz (central filled circle) or an AT20G visibility ratio at 20\,GHz (filled annulus). Sources detected in the 20\,GHz survey tend to have their flux entirely localised to a $\leq0.5$\,arcsec region at that frequency, while sources detected in IPS have anywhere from $30$--$100$ per cent of their flux contained within a single, compact component. If a source was detected in both the MWA-IPS survey and the AT20G survey, this is indicated by a datapoint with both a shaded inner circle and annulus.}
    \label{fig:compactness-subarcsec}
\end{figure}

At higher frequencies, there is another tool used by \citet{Chhetri2013TheCatalogue} to probe comparable sub-arcsecond structure called the visibility ratio. In that work, the ratio of complex visibilities from long and short baselines was used to measure the relative flux contained within 0.5\,arcsecond structure at 20\,GHz, by making use of the original visibility data from the Australia Telescope Compact Array (ATCA) 20 GHz survey \citep[AT20G;][]{Murphy2010TheCatalogue}. Of our sample, 21 sources are contained within that catalogue, and all of these have visibility ratios $>0.9$, indicating more than 90 per cent of their flux is contained within a sub-arcsecond component. Since both this visibility ratio and the NSI from IPS probe comparable angular scales with metrics spanning the same range, we have combined their values in Figure~\ref{fig:compactness-subarcsec}, where the central fill-colour of points indicates their NSI, and the annulus colour is the visibility ratio for the same source. 

Evidently, there are selection effects at play here. Even a strong, Jy-level source peaking at 20\,GHz in the observers' frame will almost certainly fall well below the sensitivity threshold of the IPS survey at 162\,MHz, and vice versa. Indeed, there are only 3 sources which are detected and measurably compact at both frequencies. Therefore by combining data from each we can more uniformly study the structure of our full PS sample. Even so, the 162\,MHz IPS measurement is more likely to probe the absorbed component below the spectral turnover in our sample, and the visibility ratio will probe the converse, since the subset of our sample detected in these surveys peaks between $\sim100\,\text{MHz}$--$20\,$GHz. Any extension in a PS source should be most visible below the spectral peak, where the absorbed emission traces the very edges of the jet structure. It is therefore not surprising that the NSI of many objects peaking $>200\,$MHz is somewhat lower than 1. Although a large fraction of their flux still comes from a compact core, some non-negligible amount may be traced to more extended structure. By contrast, the visibility ratio is probing the optically thin emission above the peak, which, based on the data here, is coming almost entirely from a sub-arcsecond component in those sources detected at 20\,GHz.

To put these values back into the context of source size, in our $\Lambda$CDM cosmology, both the NSI and visibility ratio probe scales spanning $0.3$--$2.5$\,kpc at redshifts $0.1 < z < 5$, which neatly encapsulates the range of sizes PS sources are thought to span. What Figure~\ref{fig:compactness-subarcsec} demonstrates then, is that where our sources are detected with the 20\,GHz visibility ratio, they do indeed appear to have a linear extent less than $\sim2\,$kpc. Those detected by the 162\,MHz IPS technique may have some extended structure at lower frequencies, possibly indicative of older jet activity, but nevertheless a significant fraction of their flux still comes from a compact, parsec to kiloparsec-scale component. Further expansion upon the IPS work of \citet{Morgan2018IPS-method, Chhetri2018InterplanetaryFrequencies} and the visibility ratios of \citet{Chhetri2013TheCatalogue}, increasing the sensitivity and sky coverage of both samples would offer an invaluable contribution to the study of sub-arcsecond source structure, allowing us to probe structures too small for typical continuum imaging, and yet too large for usual VLBI baselines.

\subsection{Radio variability}\label{sec:variability}

Alongside its linear extent, another key feature of a PS source is its non-varying radio spectrum. The exact threshold for ``low variability'' in a broadband SED is not well defined in the literature, but it is accepted that, at the very least, the overall spectral shape of a PS AGN should not appreciably change on timescales less than a few years \citep{EdwardsTingay2004, Hancock2010}. By contrast, \citet{Ross2021} showed that the variability of sources peaking at or below 250\,MHz is greater than in their higher-peaked counterparts, but that many of these sources were identified with known blazars. 

\begin{figure}
    \centering
    \includegraphics[trim={0cm 0cm 0cm 0cm}, width=0.45\textwidth]{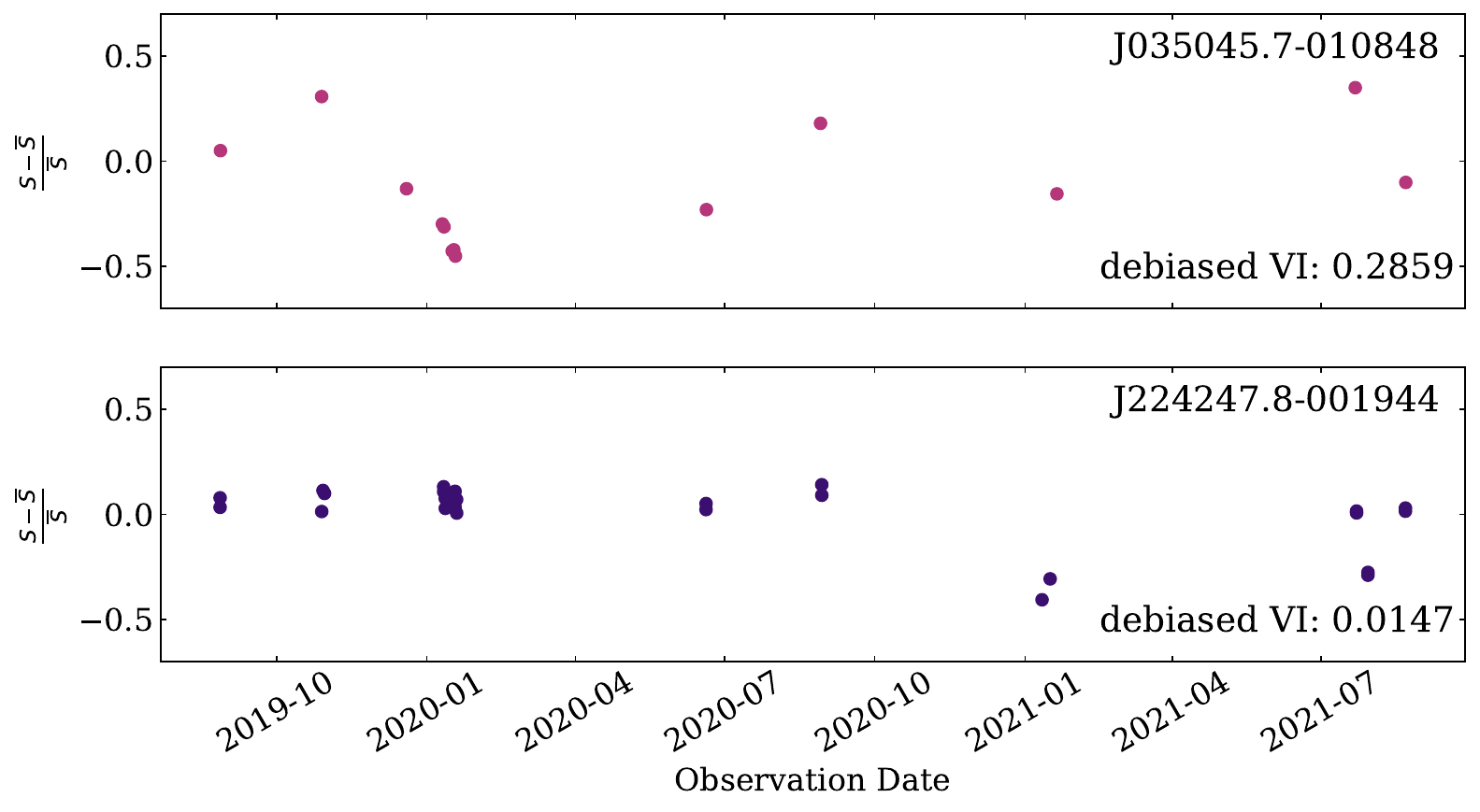}
    \caption{Two example radio lightcurves drawn from amongst our PS sample, one with a debiased variability index indicative of variability (top panel), and one that is essentially non-varying to within the sensitivity of the VAST lightcurve (bottom panel).}
    \label{fig:lightcurves}
\end{figure}

We here consider the variability of our sample at our selection frequency ($888$\,MHz) by making use of another ASKAP survey, the Variables and Slow Transients Survey (VAST; \citealt{Murphy2013VAST}). The VAST Pilot surveys, which we use here, comprise two phases, the first of which is made up of $162\times12$ minute observations made between August 2019 and August 2020, with an observing cadence of between 1 day and 8 months \citep{MurphyVASTPilot}. Phase II data extends these observations until November 2021. Between both Pilot Surveys, the entirety of the Stripe 82 field is covered by at least nine unique epochs; the radio lightcurves of our sample will therefore be sufficiently well sampled for simple variability calculations. Since the VAST Pilot observations span two years, studying PS variability with them is a valuable test to see whether our sample behaves more like the targeted samples of \citet{EdwardsTingay2004}, \citet{Jauncey2003PSVariability} and others which are largely non-varying but typically monitored at $\geq1.4\,$GHz, or like the low-frequency sample of \citet{Ross2021}. Two example lightcurves, one showing significant variability, and one non-varying, are shown in Figure~\ref{fig:lightcurves}.

To measure the radio variability of our sources, we follow \citet{Barvainis2005} and \citet{Sadler2006} in using the debiased Variability Index (VI):
\begin{equation}\label{eqn:debiasedVI}
    VI = \dfrac{100}{\langle S\rangle}\sqrt{\dfrac{\sum [S_i - \langle S\rangle]^2 - \sum \sigma_i^2}{N}}
\end{equation}
which is calculated on a source-by-source basis and essentially measures a per cent variation in the source flux density. Here, $S_i$ is an individual flux-density measurement, $\sigma_i$ is its associated uncertainty, $N$ is the number of data points in the lightcurve, and $\langle S\rangle$ is the mean flux density across the lightcurve. Since this statistic takes into account the uncertainty on individual datapoints and it does not specify a time-cadence for observations, it can be used to meaningfully compare variability measures across our entire sample even though our sources fall into several different VAST pointings, and thus were observed at slightly different times and cadences. It will also be minimally impacted by any variations in field calibration.  

It is possible for the numerator within the square root of VI to return a negative value if the uncertainties are large enough (an issue particularly for low signal-to-noise sources). In such cases, we follow \citet{Sadler2006} in first setting the VI for these sources to be negative, and then using maximally-negative VI from across our sample to define the sensitivity limit of our data. In this case, the value was found to be 6.8 per cent, and so we consider all VIs below this value as upper limits, indicating a variability $<6.8$ per cent on the timescales of the VAST pilot survey. This value was calculated using the peak flux density of sources in each epoch (rather than the integrated flux density) to minimise the effect of the source fitting algorithm.

\begin{figure}
    \centering
    \includegraphics[trim={4.75cm 2cm 4cm 0cm}, width=0.5\textwidth]{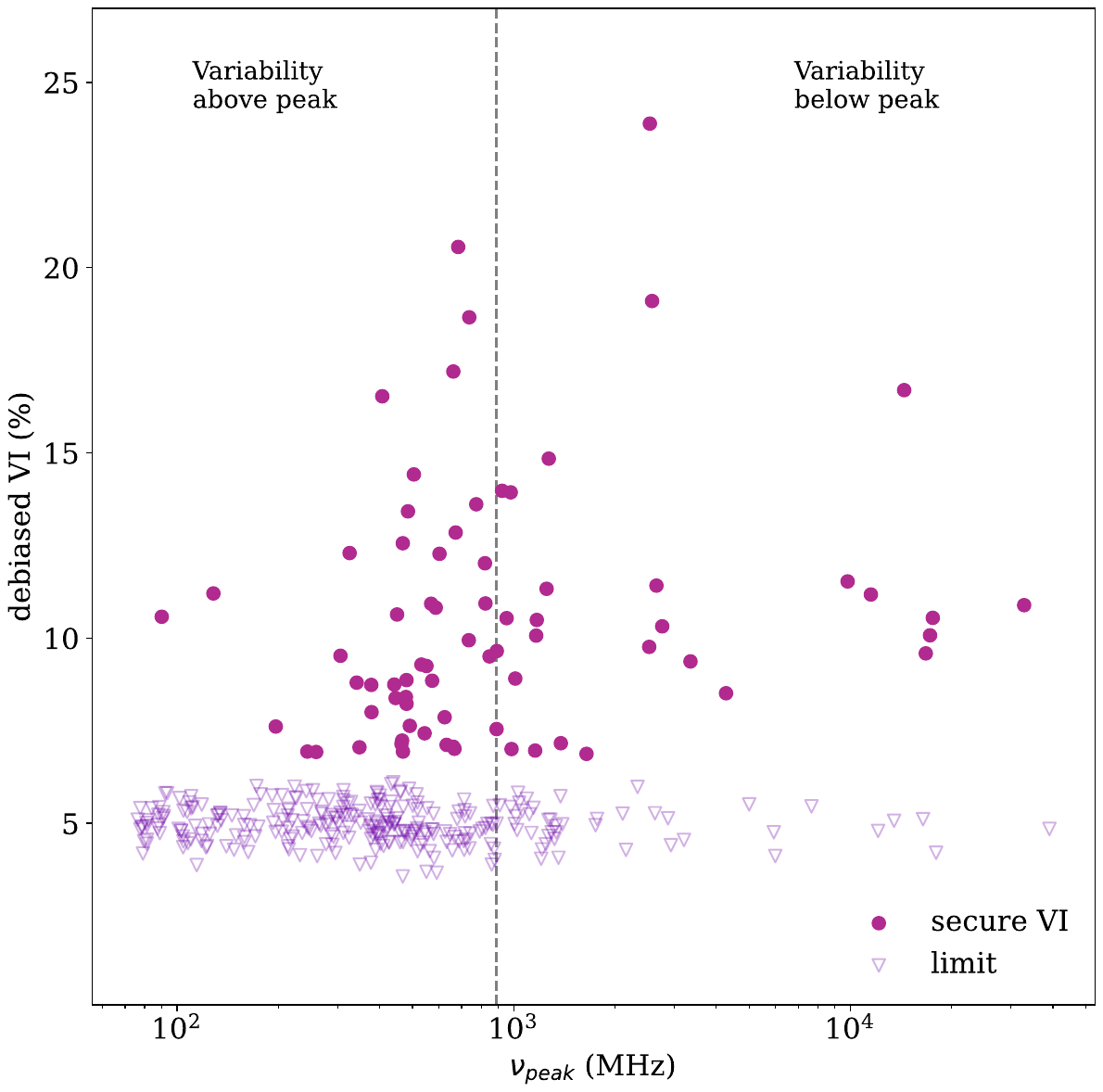}
    \caption{The debiased Variability Index of our sources as a function of peak frequency $\nu{_\text{peak}}$. Sources for which the measured VI falls below the sensitivity of the VAST Pilot lightcurves are indicated by open arrows, filled circles are those with a secure VI measurement. The vertical dashed line marks the observing frequency of VAST, which divides the plot into sources where we have effectively measured variability above the peak (left) and below the peak (right). Sources for which we have only upper limits on VI have been scattered about a mean value of 5 per cent for ease of viewing.}
    \label{fig:debiased_VI}
\end{figure}

The variability of our PS sample is summarised in Figure~\ref{fig:debiased_VI}, where the VI is plotted against the peak frequency of the sources, as obtained from \textsc{RadioSED}. It is clear that the vast majority (73 per cent or 259 sources) have only an upper limit on their VI, consistent with a very low degree of variability. Of the remaining 96 sources, none have a VI above 25 per-cent. We note that 4 sources from amongst our \fullPScount\, set do not have a measured VI from the VAST Pilot data. This is because in source finding, they were divided into multiple components in at least some of the VAST images (often due to strong sidelobes confusing the sourcefinding algorithm), so it would not be meaningful to compare their measured flux densities across different epochs.

However, we must now return to the question of what constitutes ``a low degree of variability'', as 25 per cent seems on first inspection relatively high. Indeed, at 4.8\,GHz \citet{EdwardsTingay2004} found 18/20 of their GPS sources had a variability index less than 8 per cent (the median of their larger, parent sample), and at 20\,GHz \citet{Hancock2010} found an average variability of 10--12 per cent for their sample of 21 candidate GPS sources. Our sample median is in line with these previous results, coming to 8 per cent for just those sources above the sensitivity limit. Yet the fact remains that there is a significant tail to the distribution of VI in our sources, which warrants further consideration.

\begin{figure}
    \centering
    \includegraphics[trim={0cm 1.5cm 0cm 0cm}, width=0.5\textwidth]{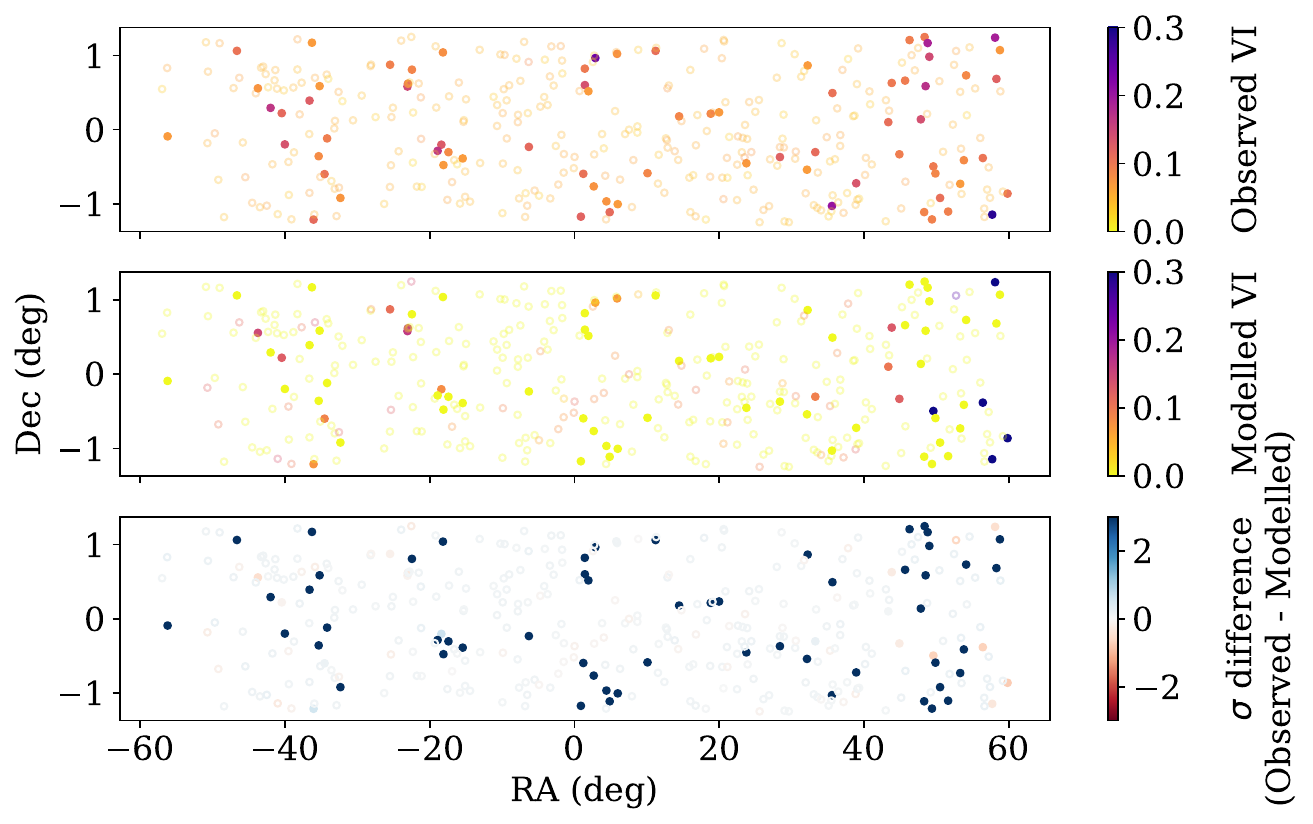}
    \caption{Observed debiased VI (top panel) compared to that predicted by the \citet{Hancock2019RISS} model (middle) assuming a source size of either $\theta_\text{src} = 0.5$\,arcsec or $\theta_\text{src} = 1.5\times10^{-7}$\,deg if a source is included in \citet{Petrov2025}. The bottom panel shows the difference between the observed and modelled values. Unfilled circles are those sources where the observed debiased VI is below our sensitivity limit, and so comparison with the model is not meaningful.}
    \label{fig:hancock}
\end{figure}

It has been previously found through VLBI observations that some PS sources exhibit extrinsic variability due to Interstellar Scintillation \citep[ISS;][]{Jauncey2003PSVariability}. To test whether this may be at play here we adopt the model presented in \citet{Hancock2019RISS}, which characterises how Refractive Interstellar Scintillation (RISS) can manifest as radio frequency variability in compact sources -- including extragalactic sources -- on timescales of months to years. The model assumes a scattering disk (source size) of approximately $\theta_\text{src} = 1$\,mas at the galactic latitude of the Stripe 82 field, which is used to produce a modulation index ($m_p$) at the location of every input source. However, it is unlikely that the majority of flux in our PS sources is concentrated in an area this small. Accordingly, we scale the output $m_p$ from the model by a factor:
\begin{equation}
    m_e = \left(\dfrac{\theta_\text{scat}}{\theta_\text{src}}\right)^{\frac{7}{6}}m_p
\end{equation}
following equations 12 and 17 from \citet{Hancock2019RISS}. The question now reduces to a question of the angular size of our sources. Conservatively, we assume an angular size of $\theta_\text{src}=0.5$\,arcsec, which is the approximate scale to which the IPS measurements were sensitive in Section~\ref{sec:structure-parsec}. For an additional constraint, we crossmatch our sample with the `Astrogeo' VLBI catalogue of \citet{Petrov2025}, which is an extensive sample of VLBI sources observed at 2.2\,GHz and up. This provides us with VLBI measurements for 74 of our sources, of which 19 have VIs above the sensitivity threshold of our data. The shortest baselines used in the Astrogeo catalogue are somewhere below $1\,000$\,km, if we take this and the lowest observing frequency (2.2\,GHz) as reference values, a source detected in that catalogue will have some significant flux on scales $\theta \sim 1.5\times10^{-7}$\,deg. Again, making the assumption that flux is still contained on this angular scale at the VAST frequency (888\,MHz), we can use this as a new value for $\theta_\text{src}$ in those sources appearing in that catalogue. This is what is shown in Figure~\ref{fig:hancock}, where the filled circles in the top panel correspond directly to the filled circles in Figure~\ref{fig:debiased_VI} as sources with a well-constrained VI. Sources that appear as unfilled circles in all three panels are those below our sensitivity limit in VI, so no useful comparison to the models can be made.

Clearly, the majority of sources with well-constrained VI have variability which is above the level of RISS with the source sizes assumed, which would suggest an intrinsic source of variability. Crucially though, over half (12/19) of the sources which appear in Astrogeo have a modelled VI that is larger than that measured here, suggesting that the radio variability detected in VAST could be consistent with external effects caused by a scattering screen. This is not the first time such a phenomenon has been observed towards PS sources, as \citet{Jauncey2003PSVariability} identified at least one of their GPS candidates which exhibited variability directly attributed to ISS. Since the Astrogeo catalogue is not complete, in the sense that it does not contain VLBI measurements for every known source in any region of sky, it is entirely possible that more of our PS sample have a significant fraction of their flux contained within small angular scales, and that some of the variability seen here may indeed be due to RISS. 

To say more on the variability of these sources is beyond the scope of this paper, so we end this section simply by reiterating that our PS sample on the whole does show a low degree of variability, with a median of just 8 per cent amongst those sources above the sensitivity limit in VI. This is in line with the higher-frequency findings of \citet{Jauncey2003PSVariability, EdwardsTingay2004} and \citet{Hancock2010}, and none of our sources show extremely high variability indicative of blazar-like activity. Though some of our sources do exhibit variability as high as 25 per cent, some portion of this may be due to RISS, especially given that a number of sources detected in the VLBI Astrogeo catalogue have significant flux on scales compact enough to induce a large degree of scattering. However some of this variability may of course be intrinsic to the source and caused by events such as transient shocks or secular source evolution.

\section{The hosts of Peaked Spectrum sources}\label{sec:host_optical}

In this section we outline the properties of the host galaxies for our PS sample, where these sources have been detected at optical and/or infrared wavelengths. Since the main focus of this work is on the nature of the radio sources themselves, we restrict this discussion mainly to a consideration of the cosmological distances of our sources, as well as the classification of their hosts in the broadest terms as galaxies or quasars (QSOs). We show that 97 of our sources have spectroscopically identified redshifts, and a further 69 have some constraints on their distance from photometric redshifts. The spectra and photometric colours show that our sources are fairly evenly distributed amongst galaxy-type and QSO-type hosts, while combining this data with WISE photometry reveals that many of our sources are likely distant ($z \geq 1$).

\subsection{Spectroscopy}\label{sec:spec_z}

We obtain spectroscopic redshifts for our sources by crossmatching primarily against the LARGESS \citep{Ching2017}, SDSS-DR16 \citep{Ahumada2020SDSS}, and the Dark Energy Spectroscopic Instrument Data Release 1 (DESI-DR1) \citep{DESI-DR1} catalogues. Additionally, we search the NASA/IPAC Extragalactic Database (NED) for supplemental spectroscopic redshifts, of which we obtained three. Since all of our sources are compact as outlined in Section~\ref{sec:structure-parsec}, we simply search for optical matches within a 5\,arcsecond radius across the catalogues which do not already provide a radio-optical association (SDSS-DR16, DESI-DR1 and NED).

In total, we obtain spectroscopic redshifts for 132 sources in our sample, of which 69 have QSO-like spectra, and 63 are identified as galaxies. From amongst these sources, there were 6 where DESI and SDSS classifications differed (one survey pipeline preferred a Galaxy-type spectral template, the other a QSO-type template), perhaps due to the differing fibre sizes capturing different ratios of light from the AGN and host galaxy. Manual inspection of these sources revealed unresolved point sources in the optical images, with visible components from both galaxy and AGN-driven emissions in the spectra, and in a more complex classification scheme these might be called `Seyfert 1' or `Broadline AGN' depending on their radio loudness. For simplicity here, we have included these in the `QSO' class. The spectroscopic redshifts of these sources span $0.04 < z < 3.7$, though above $z \sim 1.7$ all of our spectroscopically-identified sources are quasars, a natural consequence of the magnitude limits on optical surveys. We engage in a fuller analysis of these redshifts below, where they can be considered alongside photometrically-identified redshifts.

\subsection{Photometry}\label{sec:photo_z}

We supplement our spectroscopic redshifts with photometric redshift estimates drawn primarily from two large-area surveys which cover Stripe 82 in its entirety: the Quaia catalogue \citep{Storey-Fisher2024}, and the WISE-PS1 catalogue \citep{Beck2022}. Both of these combine photometry from the Wide-field Infrared Survey Explorer (WISE; \citealt{Wright2010}) with an optical survey, though each targets a different class of extragalactic sources. Specifically, Quaia combines unWISE and Gaia photometry to target quasar photometric redshifts, while WISE-PS1 makes use of Pan-STARRS1 photometry to target galaxies. These two datasets are therefore complementary, as they cover independent subsets of our sample. Furthermore, a comparison of the photometric and spectroscopic redshifts for those sources in our sample possessing both shows good agreement across both the Quaia and WISE-PS1 selected subsets. To these we add three photometric redshifts from data release 9 of the Legacy Survey \citep{Dey19LegacySurvey, Duncan2022}, and a further five photometric redshifts that were returned as part of our NED search above; in each case these help characterise sources which were otherwise lacking redshift information. Overall, we find 129 photometric redshifts for our sample, of which 47 (8 QSO/39 galaxy) provide constraints on sources without spectroscopic identifications.

\begin{figure}
    \centering
    \includegraphics[trim={0cm 1.5cm 0cm 0cm}, width=0.5\textwidth]{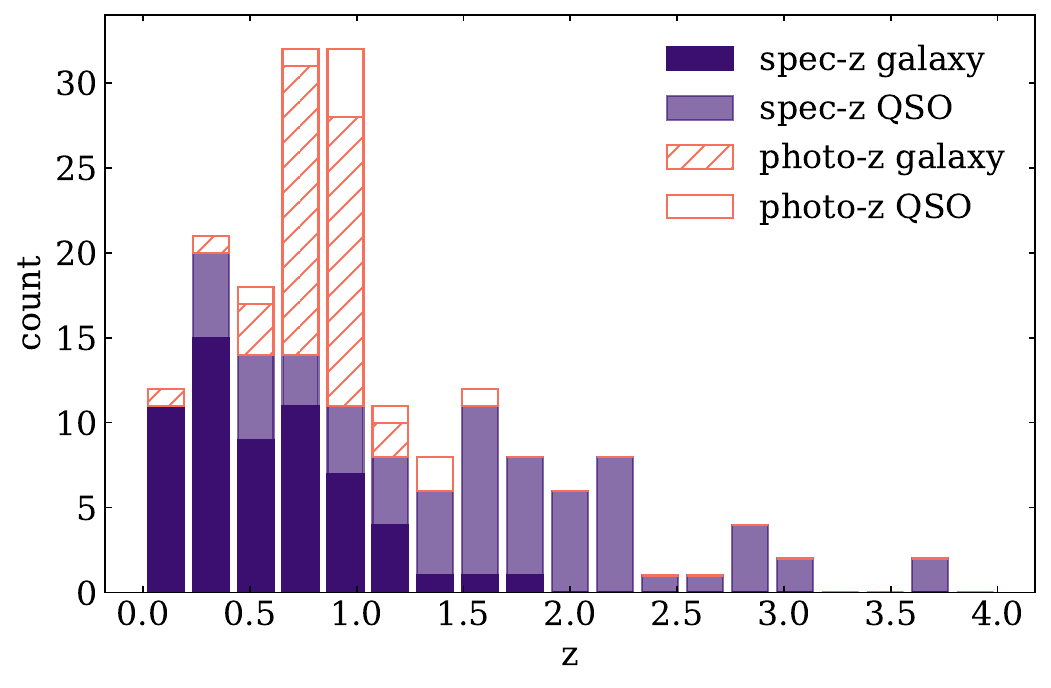}
    \caption{The distribution of spectroscopic (filled) and photometric (unfilled) redshifts for our sample, divided into galaxies (dark filled/hatched) and QSOs (light filled/unfilled). The apparent sharp declines in photometric redshifts after $z \sim 1$ and spectroscopic redshifts after $z\sim 2$ are discussed in text. }
    \label{fig:redshift_dist}
\end{figure}

 A histogram of the combined spectroscopic and photometric redshift estimates is given in Figure~\ref{fig:redshift_dist}, where it is clear that a relatively large fraction of galaxy-type photometric redshifts (hatched bars) are clustered around $z = 1$. This is likely an artefact of the WISE-PS1 catalogue used, as redshift estimates in that work are made via a machine learning approach for which the training set of spectroscopic galaxy redshifts tails off sharply at $z = 1$ (see Fig. 1 in \citet{Beck2022} for reference). Nevertheless, we keep these redshifts as a useful first-order constraint on sources that otherwise have no distance or host information (that is, are not otherwise divided into galaxy or QSO type sources).

In addition to the cut-off in photometric redshifts, we see the distribution of spectroscopic galaxy (shaded, dark) and QSO (shaded, light) redshifts fall off at $z\sim 1$ and $z\sim 2$, respectively. In the case of the galaxies, this is likely due to the underlying sensitivity limit of SDSS, while for the quasars it is a consequence of the targeting algorithm which, as of SDSS-IV, was designed to maximise completeness at $z\leq 2.2$ \citep{Myers15}. Taking this into account alongside the discussion of the photometric training sample above, there are no obvious trends in the redshift distribution of our PS sample. What is perhaps more interesting then, is a consideration of the sources that lack redshift estimates, and indeed also optical photometry. For these, we turn to an analysis at infrared wavelengths in the following section.

\subsection{WISE colours}\label{sec:wise}

For a fuller consideration of the host galaxies of our sample, we now analyse their redshift and optical classification in conjunction with IR photometry from the WISE survey. For this, we first match against the AllWISE catalogue \citep{Cutri2013}, again using a simple, radial constraint of 5\,arcseconds. This returns a total of 250 unique matches from amongst our sample, corresponding to a match rate of $\sim70$ per cent.

\subsubsection{The WISE colour-colour space}\label{sec:colour-colour}

\begin{figure}
    \centering
    \includegraphics[trim={1cm 0.5cm 0.7cm 0cm}, width=0.45\textwidth]{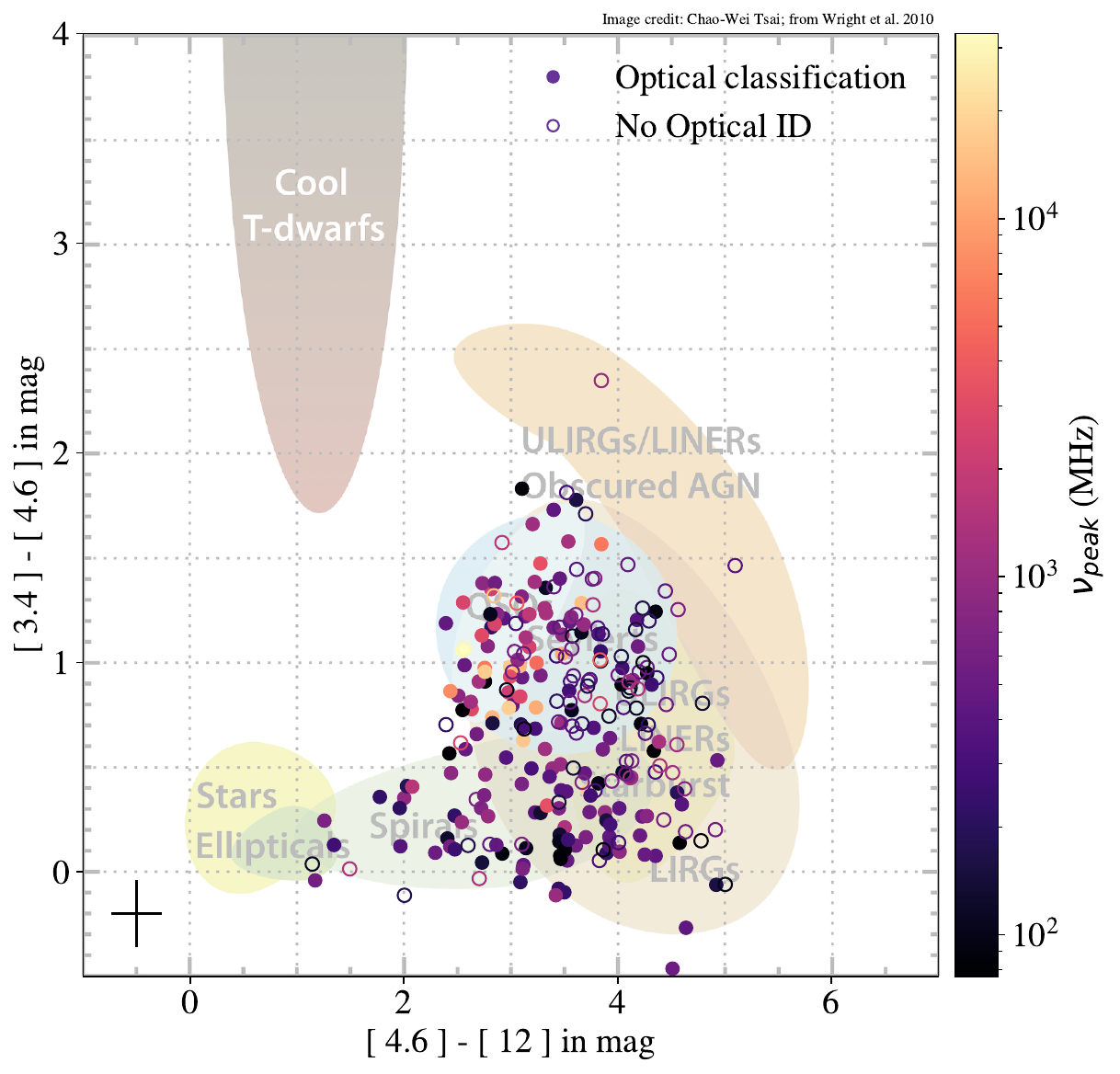}
    \caption{The distribution of our sample in the two colour WISE diagram, with the familiar classification groups from \citet{Wright2010} overlaid for comparison. Filled points have an optical classification from either spectroscopy or the photometric catalogues we referenced in Section~\ref{sec:photo_z}, unfilled points do not. Points are coloured by the peak frequency of their broadband radio SED, $\nu_\text{peak}$. The average uncertainty on these colours is indicated by the errorbars in the bottom left of the plot.}
    \label{fig:wise_colour_colour}

\end{figure}

The locus of our sources in the WISE colour-colour plot is shown in Figure~\ref{fig:wise_colour_colour}, with points coloured according to the turnover frequency in their radio SED. There are two striking features about the IR properties of these sources which may in fact be linked; the distribution of peak frequencies, and the lack of Early-type hosts. 

Beginning with the first of these, we see that the sources with the highest frequency spectral turnover are most closely clustered around the region typically inhabited by QSOs/Seyferts, with $\nu_\text{peak}$ falling as the WISE colours of sources deviate further from this space. This is quite interesting, as even without accounting for the redshifting of the peak frequency, the QSOs in our sample would seem to have smaller jets than their galaxy-type counterparts, recalling that on the whole, the QSOs are more distant in our sample than the galaxies (meaning their peaks would be redshifted to comparatively lower frequencies than galaxy-type sources; see Figure~\ref{fig:redshift_dist}). Furthermore, where available, the optical classifications of our sample support this finding, with the QSO-type objects falling comfortably within the QSO/Seyfert space of the WISE colour-colour plot. This reinforces the notion that the sources with the highest frequency peaks are indeed hosted by QSOs. The absolute $i-$band magnitudes of these sources ($M_i$) are also indicative of QSO activity; all of our sources possessing an optical redshift and WISE colour $W1 - W2 > 0.5$ are brighter than $M_i < -22\,\text{mag}$, which is the tail end of the $i-$band absolute magnitudes in the SDSS-DR16 Quasar Catalogue \citep{Lyke2020TheRelease}. This may suggest that there is something unique about the environment of QSOs which tends to trigger the very smallest radio jets, whereas sources with galaxy-type optical hosts are less likely to do so, and indeed galaxy-type objects with radio peaks above a few gigahertz were not detected in our sample.  In support of this interpretation, we note that \citet{Hancock2009} found the majority of PS sources identified at gigahertz frequencies in AT20G were quasar-type, while the recent work of \citet{Nyland2020} looked specifically for changing-look quasars in the Very Large Array Sky Survey (VLASS) at 3\,GHz, and identified several with radio SEDs indicative of recent triggering (i.e. with high frequency peaks), suggesting a selection-philosophy broadly consistent with our interpretation here. 

On the subject of the broader distribution of our sample in the WISE colour-colour space, we note that only $\sim3$ per cent (8/267) of our PS sources detected in all four WISE bands would be considered `Wise Early-type' ($W2 - W3 < 2$), while some 47 per cent of optically-identified sources are hosted by QSOs. This latter fraction is similar to the proportion of QSOs in one of the first canonical samples presented in \citet{ODea1998}, but the low fraction of Early-type galaxies is worth a further moment of consideration. It is lower than the fraction of Early-type galaxies in the young radio galaxy sample of \citet{Kosmaczewski2020}, though not by much considering the small size of that sample ($5/29 \approx 17$ per cent are Early-type in that work), and the additional selection criteria imposed there (X-ray detection, spectroscopically-constrained redshift, constraint on linear size of the radio jets). What is perhaps more interesting is a comparison with the ``Fanarof-Riley Type 0'' (FR0) sample of \citet{Sadler2016}, of which almost 67 per cent  exhibit a low star formation rate consistent with Early-type host galaxies. That sample was constructed based on the relatively sparse frequency-coverage available at the time. Therefore, the authors did not require their sources to exhibit broadband spectral turnover, only compactness at 1\,GHz in addition to the 20\,GHz selection frequency, making it a sample of \textit{candidate} PS sources. Nevertheless, the only additional selection criterion imposed was a crossmatch with an optical counterpart, which makes it a useful point of comparison against our work. The discrepancy between the mid-IR colours of that sample and ours is therefore striking, and suggests we are perhaps probing some subset of that larger sample of compact objects, which does not include sources with Early-type hosts. Indeed, the WISE colours of our sample are more akin to those of the extended, lobe-dominated (FR-II) radio galaxy sample presented in \citet{vanvelzen2015} --- discussed further with reference to PS sources in \citet{Kosmaczewski2020}. This once more raises the question, discussed in \citet{Sadler2016}, as to the connection between PS sources specifically, and the broader class of compact radio objects; how these two different classification methodologies overlap, and what can be said about how each type of radio source will evolve (or not) into more extended structures as it ages, a discussion recently revived in the series of papers beginning with \citet{Kiehlmann2024}.

\subsubsection{WISE photometry and approximate source distances}

\begin{figure*}
    \centering
    \includegraphics[trim={0cm 5.5cm 6cm 0cm}, width=0.95\textwidth]{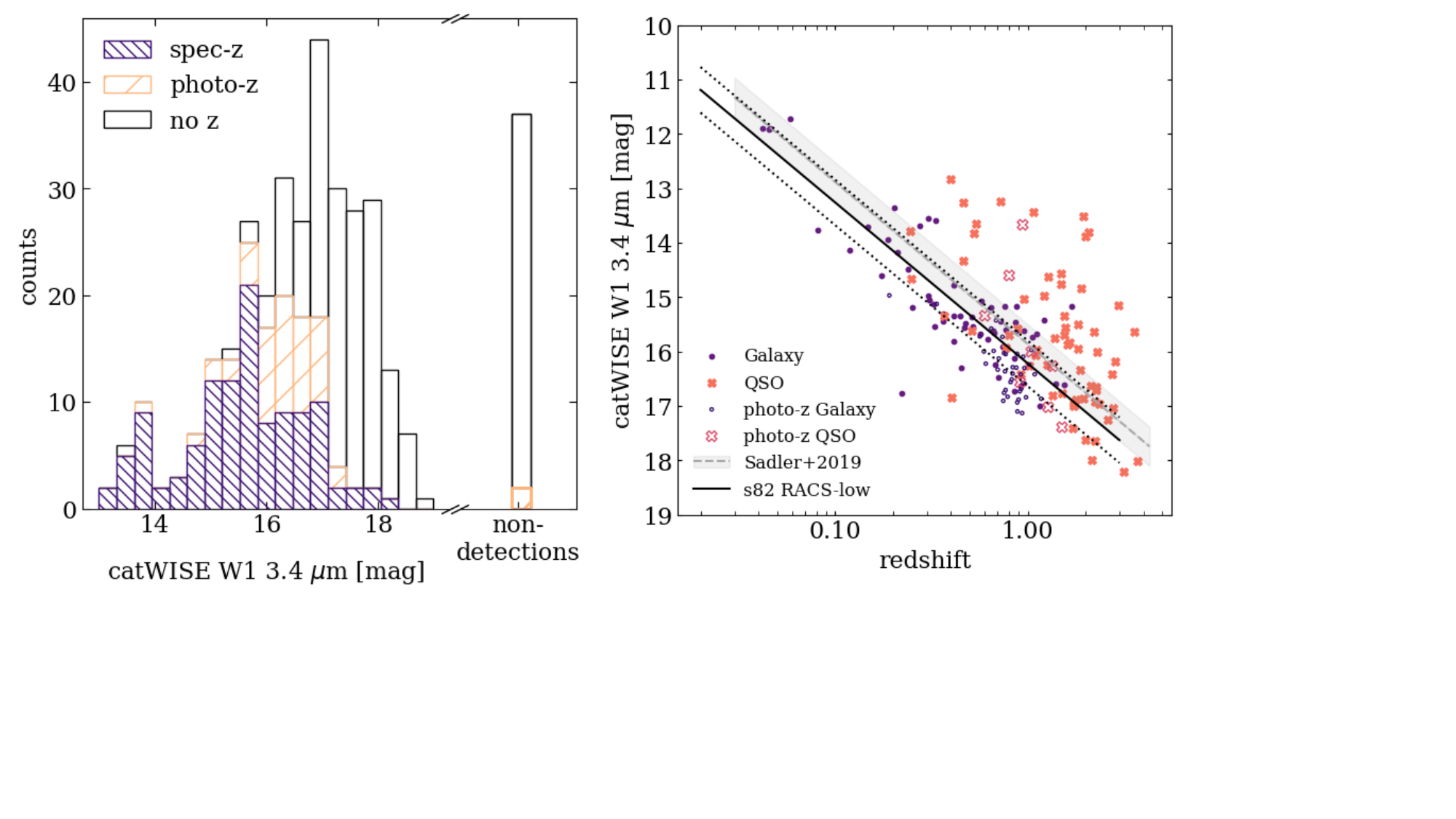}
    \caption{Left: The distribution of spectroscopic (densely hatched, purple) and photometric (sparsely hatched, orange) redshifts for our sample as a function of WISE W1 magnitude, as well as sources without any redshift constraint (unfilled). Some 38 of the sample have neither WISE photometry nor a redshift estimate, which are candidate high-redshift sources as discussed in-text. Right: Spectroscopic (filled points) and photometric (unfilled points) redshifts for our sample as a function of WISE W1 magnitude, divided into QSO-type (crosses) and Galaxy-type (circles) sources. Dashed grey line is the linear fit from \citet{Sadler2019}, with shading indicating the median scatter in that sample.  Solid black is fit to a full sample of RACS-low radio sources from Stripe 82, with black dotted lines indicating the median scatter in this sample. Some 38 of the sample have neither WISE photometry nor a redshift estimate, which are candidate high-redshift sources as discussed in-text.}
    \label{fig:wise_nondets_pair}
\end{figure*}

We see also in Figure~\ref{fig:wise_colour_colour} a certain clustering of the the sources without optical redshifts, either spectroscopic or photometric. Again, these appear to be centered around the typical locus of QSOs/Seyferts and starburst galaxies, with just a very few sources possessing WISE colours indicative of an elliptical, or an obscured AGN. This may suggest many of our sources without redshift information are high-$z$ QSOs. To explore this further, we redo the IR crossmatch with the catWISE 2020 catalogue \citep{Marocco2021} to obtain greater W1 completeness for our sample (catWISE2020 is 99 per cent complete for $W1 < 15.1\,\text{mag}$, whereas AllWISE is only 95 per cent complete at these magnitudes). Since catWISE2020 only contains information for the first two WISE bands, we perform a simpler analysis presented in Figure~\ref{fig:wise_nondets_pair}. In the left-hand figure, we show the distribution of sources as a function of W1, broken down into those with spectroscopic redshifts (right-hatched, navy bars), photometric redshifts (left-hatched, orange bars) and without redshift information (unfilled bars). It is clear that redshift completeness is a strong function of $W1$ magnitude, with only 9 per cent of sources (14/154) fainter than W1 $= 17$\,mag possessing redshift information.

The WISE magnitudes of our sources can tell us something about their distances, even in the absence of redshift information, in a variation on the familiar $K-z$ relation. Although this relation only holds for galaxy-type hosts (it relies on a simple scaling of flux from star formation with redshift), it can also be used as a lower-limit on QSO-type hosts, where the optical-IR light is dominated by the central AGN but nevertheless will have some contribution from starlight. \citet{LaMassa13} discussed this in relation to X-ray sources within Stripe 82, noting that sources undetected in WISE or optical photometry could be either highly dust-obscured, or candidate high-redshift sources. Closer to our frequencies of interest, \citet{Sadler2019} showed that the W1 magnitudes of a number of radio galaxies detected at 162\,MHz with the MWA could be fit by the following linear equation:
\begin{equation}\label{eqn:sadler}
    W1 = 15.860 + 2.976\log(z)
\end{equation}

We plot the W1 magnitudes of our sample as a function of redshift in the right-hand panel of Figure~\ref{fig:wise_nondets_pair}, where filled markers indicate spectrosocpic redshifts, and unfilled photometric. Equation~\ref{eqn:sadler} is shown as a dashed grey line, with grey shading either side indicating an offest of $\pm0.36$\,mag, the median scatter in their sample. On top of this we have added a solid black line indicating a similar linear fit, but this time to all RACS-low point sources within Stripe 82 (i.e. sources with 1 Gaussian component in the radio catalogue) which are brighter than 10\,mJy, have a spectroscopic, galaxy-type redshift from SDSS-DR16 and a catWISE counterpart within 3 arcseconds of the radio position. This fit has the form: 
\begin{equation}\label{eqn:s82_w1}
    W1 = 16.218 + 2.958\log(z)
\end{equation}
and the median scatter in the sources about this fit is $\pm0.42$\,mag, as indicated by the dotted lines. This new fit corresponds fairly closely to the one derived previously for compact radio galaxies bright at 162\,MHz, just shifted down by approximately 1$-\sigma$, which suggests that radio galaxies selected at our chosen frequency (888\,MHz) and sensitivty ($S \geq 10$\,mJy) overall reside in slightly less luminous hosts than those in the \citet{Sadler2019} sample, which were selected to be highly luminous at radio wavelengths, and are also likely to have a strong non-thermal contribution to their WISE W1 magnitude. As is to be expected, many of the QSOs in our sample lie above both relations, since the central AGN can provide a strong excess of optical-IR emission over that provided by extended star formation. However, the galaxy-type sources in our sample lie largely below the \citet{Sadler2019} fit, and this offset still persists when considering our fit to a radio-matched sample, albeit to a much lesser degree, as the median offset of our PS sample is 0.23\,mag fainter than the simple linear fit. From this we conclude that our sample of PS sources reside in somewhat less luminous hosts than the bright, strongly scintillating sources of \citet{Sadler2019}, and also in slightly less luminous hosts than the majority of other radio galaxies in the field.

Even so, we can use our linear fit in Equation~\ref{eqn:s82_w1} to obtain a first estimate for the distances of the objects for which no redshift is available. As was stated in \citet{Sadler2019}, we caution that this is only an estimate of the lower-limit on source redshifts, as the QSOs typically have an additional flux contribution from their core as discussed above, and are thus not well described by the fit we performed. Nevertheless, as stated above the vast majority of our sources without spectro-photometric information have catWISE magnitudes W1 $> 17$\,mag, which would put them at redshifts $z\geq1.8$ from our linear fit (or $z\geq1.3$ if we were to take the lower bound, recognising that our PS sources tend to fall slightly below the fit), while the 37 sources un-detected in W1 could be as distant as $z\geq 3$ (using the 90 per cent completeness limit of catWISE2020, corresponding to 17.7\,mag). 

In summary, we obtain spectro-photometric redshift estimates for 184 of our \fullPScount\, sources, divided fairly evenly between QSO-type (79/184) and galaxy-type (105/184) hosts, and largely restricted to $z\leq 1$. The WISE colours of these sources show many to be dominated by AGN emission at IR wavelengths, and we have comparatively few sources with early-type host galaxies. Following the method of \citet{Sadler2019}, we find from the relation between W1$-z$ that many of our PS sources are likely distant, perhaps as high as a redshift of $z \sim 3$, which makes this an interesting population with which to study the cosmological evolution of compact radio jets.

\section{The population of PS sources}\label{sec:PS-population}

We consider here the radio properties of our PS sample in the larger context of radio populations. We show that the luminosity distribution of our sources is typical of previous PS samples, though we are lacking some of the highest luminosity sources from earlier works. We finally consider briefly the nature of AGN environments conducive to jet launching.

\subsection{Radio Luminosity}\label{sec:radio_luminosity}
We calculate the 5\,GHz radio luminosity of the 184 PS sources with redshift estimates as:
\begin{equation}
    L_{5\text{GHz}} = \dfrac{4\pi D_L^2 S_{5\text{, rest}}}{1+z}
\end{equation}
where $D_L$ is the luminosity distance in metres, and $z$ is the redshift of the source. We note here that instead of performing a $k-$correction according to the spectral index of the source, $\alpha$, we have simply sampled the SED at a frequency corresponding to 5\,GHz in the source restframe ($S_{5\text{, rest}}$). A frequency of 5\,GHz was chosen for purely historical reasons, to allow for easy comparison with samples from the literature. 

\begin{figure}
    \centering
    \includegraphics[trim={0cm 1.1cm -0.5cm 0cm}, width=0.5\textwidth]{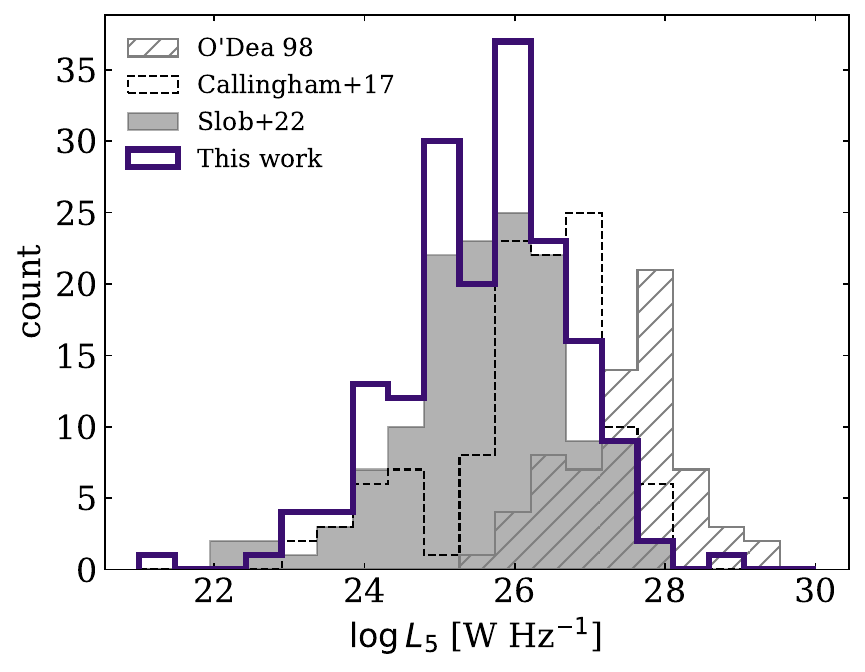}
    \caption{5\,GHz radio luminosity distribution of our sample compared to relevant ones from the literature.}
    \label{fig:luminosity_dist}
\end{figure}

The luminosity distribution of our sample is shown in Figure~\ref{fig:luminosity_dist}, along with a few key PS samples from previous work. Firstly, we note that, from amongst our full \fullPScount -source sample, the number for which we can calculate luminosities is directly comparable to the other samples highlighted here. This is despite the fact that this sample focuses on a much smaller area of sky; \citet{Callingham2017} covered the entire Southern equatorial sky, \citet{Slob2022} sampled the northern half, whereas Stripe 82 is only 300\,deg$^2$. This is likely due to both the deep optical coverage in our chosen field, but also the fact that we are maximising the use of legacy datasets to extend our search down to fainter radio flux densities. Therefore, a judicious use of legacy radio frequency catalogues across larger areas of sky is likely to increase the sample of PS sources for which we can constrain distances by several orders of magnitude.

Beyond comparisons of sample size alone, we note that the luminosity distribution of our sample most closely resembles those more recent distributions shown in Figure~\ref{fig:luminosity_dist}. As in \citet{Callingham2017}, we too lack the highest luminosity sources above $L_5 = 10^{28}\,\text{W\,Hz}^{-1}$. However, where in that work the lack of high luminosity sources from a large-area search was interpreted as a ``convolution of source evolution and redshift'', here it could be simply that our on-sky area is not large enough to see the very rarest and brightest radio galaxies. Crucially, as was seen in the \citet{Slob2022} sample, we are now starting to probe the faint luminosity end of the radio AGN population, though we are still about two orders of magnitude away from the faintest end of the local radio luminosity function where star forming galaxies begin to dominate \citep{Mauch2006}. It is very likely that future samples combining legacy radio surveys over larger sky areas than this one will greatly increase the number of such sources, and our capacity to study them will only grow with the greater sensitivity of upcoming surveys from the next generation of radio instruments.

\subsection{The significance of Galaxy-type and QSO-type hosts}\label{sec:gal_qso_discussion}

\begin{figure}
    \centering
    \includegraphics[trim={0.1cm 1.1cm -0.5cm 0cm}, width=0.5\textwidth]{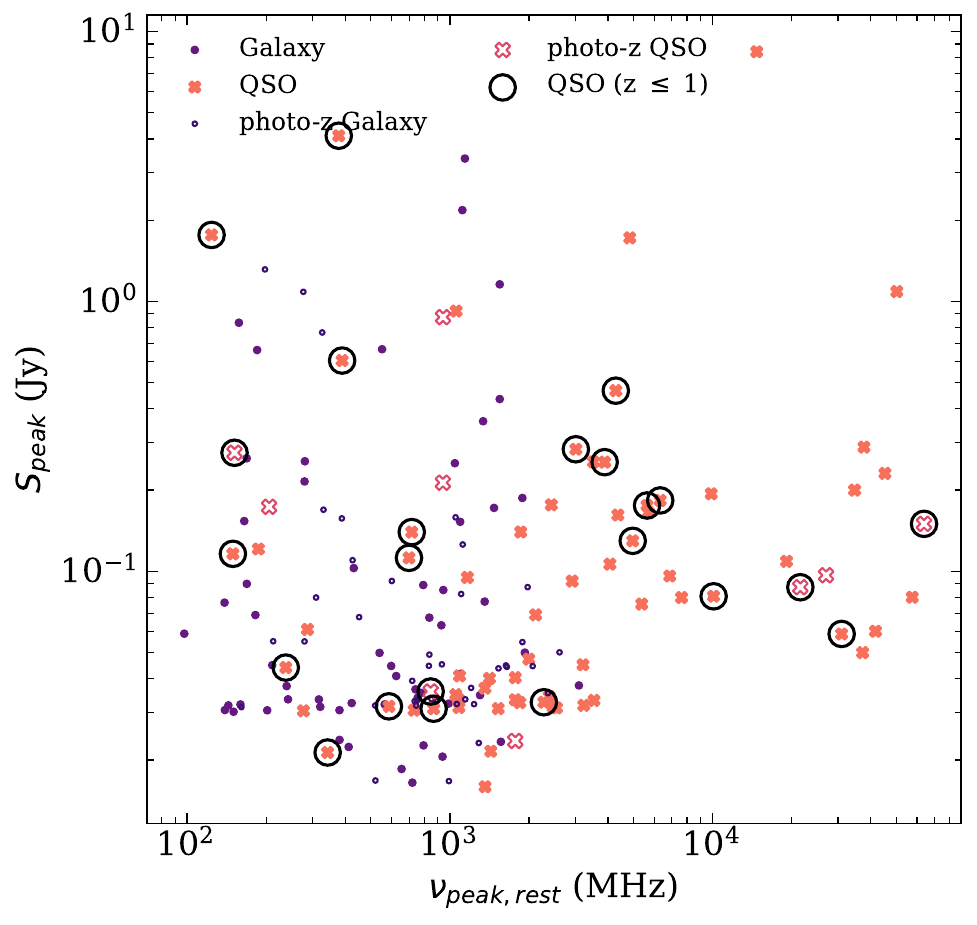}
    \caption{The distribution of peak flux densities ($S_{\text{peak}}$) as a function of $\nu_{\text{peak,rest}}$, the peak frequencies of our sample in the source rest-frame. As in Figure~\ref{fig:wise_nondets_pair} panel b, marker shapes indicate optical identification with a galaxy (circle) or QSO (cross), with filled points having a spectroscopic identification, and unfilled points photometric classification. QSOs below a redshift of 1 are encircled with a larger, black marker.}
    \label{fig:peakfreq_peakflux}
\end{figure}

In performing our analysis on this new sample of PS sources, we have placed particular emphasis on the optical identification of the hosts as either galaxies or QSOs. It is well known that jet-driven or `radio-mode' AGN emission is typically associated with radiatively-inefficient accretion, while radiatively-efficient QSOs exhibit strong, jet-driven radio emission in only $\sim10$ per cent of cases, and it is still not known exactly why \citep{Kellerman2016}. Furthermore, this fraction of jet-dominant QSOs does not appear to evolve with cosmic time, even as far as $z \approx 6$ \citep[][and references therein]{Keller2024}. Studying the conditions in which we see the most compact radio jets may therefore provide interesting insights into the complex interplay between AGN fuelling and feedback mechanisms across cosmic time.

In Figure~\ref{fig:peakfreq_peakflux} we present the peak flux density, $S_\text{peak}$, as a function of the peak frequency in the rest-frame of the source, $\nu_\text{peak, rest}$. Here, the unfilled markers indicate those sources with photometric constraint on redshift only, and filled points indicate spectroscopic redshifts are available. It is clear that the sources with the highest frequency peaks in our sample, corresponding to the most recent triggering, are QSO-type, while no galaxy-type PS sources are found to peak above $\sim 2$\,GHz in our sample. It is possible that this is partially a selection effect since quasars are optically brighter and can be seen out to higher redshifts. There may also be additional beaming effects at play, whereby intrinsically less luminous PS sources have their radio emission boosted in QSO-type hosts due to the approximate alignment of the jet-axis and our line of sight. However, if we restrict the sources in this plot to those with redshifts $z \lesssim 1$ (where the optical surveys we crossmatch against are sensitive to Galaxy-type hosts), there are still a number of QSO-type sources with peaks above a few gigahertz. These are indicated in the plot by the enclosing black circles.

Although not as marked, this phenomenon was apparent in the early, canonical sample of PS sources from \citet{ODea1998}, where only 20 per cent (4/18) sources peaking above 2\,GHz had galaxy-type hosts. This also is in line with the philosophy behind the search for changing-look quasars conducted by \citet{Nyland2020}, who found 14 radio sources peaking upwards of several gigahertz in the restframe, all hosted QSO-type objects, using a multi-epoch radio search. As already discussed in Section~\ref{sec:wise}, this suggests that radio jets may be preferentially triggered in QSO-type objects, and as they grow and extend beyond the medium of their host the central AGN may transition from radiatively-efficient accretion into the inefficient accretion typically associated with extended radio structures. If the radio luminosity of these jets increases as they grow up to scales of a few kiloparsecs, as suggested by the models of \citet{Shabala2008} and \citet{Young2024}, then such evolution of the hosts would be broadly consistent with the observed anti-correlation between radio-loudness and Eddington-scaled accretion rate \citep{Ishibashi2014}. A fuller analysis of radio jets on all-scales, coupled with detailed study of the hosts' optical spectra, will no doubt provide greater insight into this complex picture of jet triggering and growth, especially when coupled with insights from both idealised \citep{Borodina2025} and cosmological simulations \citep{Thomas2025}.

\section{Conclusions \& Future Outlook}\label{sec:conclusions}

In this work we have presented a new, uniform sample of sources with compact radio jets, thought to be recently-triggered radio AGNs. We have shown that our \textsc{RadioSED} method for fitting and classifying broadband radio flux density measurements is able to increase the number of known PS sources within the Stripe 82 field by more than an order of magnitude (Section~\ref{sec:sample-definition}), especially when making use of the latest generation of surveys including GLEAM-X DRII \citep{Ross2024} (Section~\ref{sec:fitting}). Estimates of the sensitivity of our method suggest that we are able to extend the completeness of our sample in both $\nu_\text{peak}$ and $S_\text{peak}$ by more than an order of magnitude compared to previous work (Section~\ref{sec:completeness-synthetic}). 

From amongst the \fullPScount\,sources identified by \textsc{RadioSED}, we recover all 11 previously known PS sources (Section~\ref{sec:published_ps}). Our sample is more compact than the general population of radio galaxies, at both the resolution of the continuum surveys used for fitting (Section~\ref{sec:large-structure}), and at the sub-arcsecond resolution from widefield IPS measurements (Section~\ref{sec:structure-parsec}). However, sources with `retriggered' SEDs tend to be less compact at low frequencies, likely due to extended, lobe-dominated emission. Fewer of our sources are compact at 162\,MHz than the PS sources studied in \citet{Chhetri2018InterplanetaryFrequencies}. The 800\,MHz variability of this sample is on the whole low, with the VAST pilot data only able to provide an upper limit on the debiased VI of most sources (Section~\ref{sec:variability}). In those sources which are variable, some of this variability may be explained by RISS, especially where there is evidence of milliarcsecond cores from \citet{Petrov2025}.

The multiwavelength properties of our sources suggest that these PS sources are fairly evenly divided amongst galaxy- and QSO-type hosts (Section~\ref{sec:host_optical}) in line with the divide amongst previous PS samples. However, where our sources have WISE colours fewer appear to belong to early-type hosts than in comparable FR0-type samples (Section~\ref{sec:wise}), reinforcing the idea that FR0 and PS type sources are drawn from distinct (but perhaps overlapping) populations. While many of our sources that are not spectro-photometrically constrained are likely distant, potentially residing at $z\sim3$ or higher (Section~\ref{sec:wise}), those sources for which distances are well constrained show a luminosity distribution in good agreement with previous samples (Section~\ref{sec:radio_luminosity}). When examining the rest-frame peak frequencies (a direct proxy for source size and hence also age), we see that the most compact sources in our sample tend to be QSO-type, which we discuss in the context of AGN jet triggering (Section~\ref{sec:gal_qso_discussion}).

Overall, the sample of PS sources presented here represents a new step in the study of radio AGN jets in their early stages post-triggering. Thanks to the wealth of radio and multiwavelength data already available, we are now at the stage where we can assemble large, statistical samples of these sources using tools like \textsc{RadioSED}, and begin to probe their properties on a population-level. Our capacity for this large-scale analysis will only grow with the next generation of surveys from the SKA \citep{Carilli2004} and ngVLA \citep{Murphy2018}, and with the rich datasets already available to us from decades of observing across dozens of instruments and continents.

\section*{Acknowledgements}

We would first like to thank our referee, Professor D.J. Saikia, for providing useful comments and additional references which greatly improved the overall clarity of the paper. This research was supported by an Australian Government Research Training Program (RTP) Scholarship, and by the Australian Research Council Centre of Excellence for All Sky Astrophysics in 3 Dimensions (ASTRO 3D), through project number CE170100013. Parts of this research were also supported 
by the Australian Research Council Centre of Excellence for Gravitational Wave Discovery (OzGrav), project number CE230100016.

This research has made use of NASA's Astrophysics Data System Bibliographic Services, the crossmatch service provided by CDS, Strasbourg, and the NASA/IPAC Extragalactic Database (NED) which is funded by the National Aeronautics and Space Administration and operated by the California Institute of Technology. Furthermore, this research makes use of data products from the Wide-field Infrared Survey Explorer, which is a joint project of the University of California, Los Angeles, and the Jet Propulsion Laboratory/California Institute of Technology, funded by the National Aeronautics and Space Administration. We also made use of Astroquery \citep{Ginsburg2019Astroquery:Python} and Astropy:\footnote{http://www.astropy.org} a community-developed core Python package and an ecosystem of tools and resources for astronomy \citep{astropy:2013, astropy:2018, astropy:2022}.

This research used data obtained with the Dark Energy Spectroscopic Instrument (DESI). DESI construction and operations is managed by the Lawrence Berkeley National Laboratory. This material is based upon work supported by the U.S. Department of Energy, Office of Science, Office of High-Energy Physics, under Contract No. DE–AC02–05CH11231, and by the National Energy Research Scientific Computing Center, a DOE Office of Science User Facility under the same contract. Additional support for DESI was provided by the U.S. National Science Foundation (NSF), Division of Astronomical Sciences under Contract No. AST-0950945 to the NSF’s National Optical-Infrared Astronomy Research Laboratory; the Science and Technology Facilities Council of the United Kingdom; the Gordon and Betty Moore Foundation; the Heising-Simons Foundation; the French Alternative Energies and Atomic Energy Commission (CEA); the National Council of Humanities, Science and Technology of Mexico (CONAHCYT); the Ministry of Science and Innovation of Spain (MICINN), and by the DESI Member Institutions: www.desi.lbl.gov/collaborating-institutions. The DESI collaboration is honored to be permitted to conduct scientific research on I’oligam Du’ag (Kitt Peak), a mountain with particular significance to the Tohono O’odham Nation. Any opinions, findings, and conclusions or recommendations expressed in this material are those of the author(s) and do not necessarily reflect the views of the U.S. National Science Foundation, the U.S. Department of Energy, or any of the listed funding agencies.

This scientific work uses data obtained from Inyarrimanha Ilgari Bundara, the CSIRO Murchison Radio-astronomy Observatory. We acknowledge the Wajarri Yamaji People as the Traditional Owners and native title holders of the Observatory site. CSIRO’s ASKAP radio telescope is part of the Australia Telescope National Facility (https://ror.org/05qajvd42). Operation of ASKAP is funded by the Australian Government with support from the National Collaborative Research Infrastructure Strategy. ASKAP uses the resources of the Pawsey Supercomputing Research Centre. Establishment of ASKAP, Inyarrimanha Ilgari Bundara, the CSIRO Murchison Radio-astronomy Observatory and the Pawsey Supercomputing Research Centre are initiatives of the Australian Government, with support from the Government of Western Australia and the Science and Industry Endowment Fund.

\section*{Data Availability}

 Archived data from ASKAP surveys, including VAST and RACS, can be obtained through the CSIRO ASKAP Science Data Archive, CASDA (http://data.csiro.au/).

 Other survey data used in SED construction and multiwavelength analysis is all readily available through available through the CDS VizieR catalogue service. 

Measurements derived from radio catalogues and not elsewhere published are available in the online version of this paper. Samples of these datasets are contained in Appendices~\ref{sec:appxA-sedparams} and \ref{sec:appxB-radioparams}.

 Any additional data is available upon reasonable request to the corresponding author.


\typeout{}
\bibliographystyle{mnras}
\bibliography{references} 




\appendix

\section{Radio SED parameters}\label{sec:appxA-sedparams}

We present here the fit parameters from RadioSED for our new sample of PS sources. The table here contains the first 14 entries, but the full catalogue is available online.

\bgroup
\def\arraystretch{1.5}%
\begin{landscape}
    \begin{table}
        \caption{The radio SED parameters for a small subset of the PS sources discussed in this work. Column 1 is the IAU designation for the source, Columns 2 and 3 are its coordinates in decimal degrees. Column 4 is the classification of the source based on its radio SED, either ``PS'' for Peaked Spectrum, or ``SPS'' for Soft Peaked Spectrum sources (a historical distinction that is discussed in Section~\ref{sec:fitting}). Columns 5 and 6 show the Best Model amongst those used by \textsc{RadioSED} and as outlined in \citetalias{Kerrison2024}, and the Bayes factor comparing this best model, and the next most likely one. Column 7 is the flux density of the broadband peak in Jy, and column 8 the corresponding frequency in MHz. Column 9 is the spectral index below this peak frequency, column 10 the spectral index above. Column 11 is the spectral index of the low frequency upturn (where this is detected in a source). Columns 12 and 13 relate to what we call the ``trough'', the point in the SED below which a low frequency upturn occurs, with Column 12 giving the flux density of this point (again in Jy), and Column 13 the frequency (in MHz). The full version of this table can be found in the online supplementary material.}
        \label{tab:full-cat}
    \begin{tabular}{l l l l l l l l l l l l l l}
    \hline
    (1) & (2) & (3) & (4) & (5) & (6) & (7) & (8) & (9) & (10) & (11) & (12) & (13)\\
    IAU Designation & RA & Dec. & SED Class & Best Model & Best Model $\log_{10} Z$ & $S_{\text{peak}}$ & $\nu_{\text{peak}}$ & $\alpha_{\text{thick}}$ & $\alpha_{\text{thin}}$ & $\alpha_{\text{retrig}}$ & $S_{\text{trough}}$ & $\nu_{\text{trough}}$  \\
    & (deg) & (deg) & & & & (Jy) & (MHz) & & & & (Jy) & (MHz) \\
    \hline
    \hline 
    J000001.6-002209 & 0.0068 & -0.3692 & SPS & Snellen & 4.36 & $0.28_{-0.001}^{+0.001}$ & $79_{-1}^{+1}$ & $3.96_{-0.03}^{+0.07}$ & $-0.23_{-0.0}^{+0.0}$ &  &  &   \\
    J000247.3+003111 & 0.6972 & 0.5198 & PS & Orienti & 32.84 & $0.03_{-0.001}^{+0.001}$ & $217_{-30}^{+27}$ & $0.86_{-0.24}^{+0.26}$ & $-1.53_{-0.17}^{+0.16}$ &  &  &  \\
    J000332.1-011026 & 0.8839 & -1.174 & PS & Orienti & 29.13 & $0.02_{-0.001}^{+0.001}$ & $821_{-51}^{+63}$ & $2.72_{-0.34}^{+0.37}$ & $-0.68_{-0.16}^{+0.16}$ &  &  &   \\
    J000452.3-003547 & 1.218 & -0.5965 & PS & Snellen & 26.78 & $0.17_{-0.004}^{+0.004}$ & $90_{-3}^{+3}$ & $2.36_{-0.45}^{+0.45}$ & $-0.64_{-0.02}^{+0.02}$ &  &  &   \\
    J000541.7+004918 & 1.4239 & 0.8217 & SPS & Snellen & 32.03 & $0.03_{-0.003}^{+0.002}$ & $735_{-121}^{+180}$ & $0.32_{-0.11}^{+0.06}$ & $-1.31_{-0.38}^{+0.32}$ &  &  &   \\
    J000544.7+003558 & 1.4364 & 0.5995 & PS & Snellen & 29.48 & $0.03_{-0.004}^{+0.002}$ & $979_{-157}^{+203}$ & $0.74_{-0.19}^{+0.1}$ & $-1.64_{-0.39}^{+0.61}$ &  &  &   \\
    J000741.3+003100 & 1.9222 & 0.5167 & SPS & Snellen & 34.91 & $0.03_{-0.003}^{+0.002}$ & $985_{-186}^{+287}$ & $0.33_{-0.12}^{+0.06}$ & $-0.95_{-0.36}^{+0.35}$ &  &  &   \\
    J001009.9+005440 & 2.5414 & 0.9112 & SPS & Snellen & 33.24 & $0.05_{-0.006}^{+0.003}$ & $1394_{-75}^{+182}$ & $0.31_{-0.08}^{+0.04}$ & $-2.0_{-0.6}^{+1.11}$ &  &  &   \\
    J001038.1-004559 & 2.6589 & -0.7665 & SPS & Retriggered & 32.43 & $0.04_{-0.005}^{+0.009}$ & $491_{-55}^{+66}$ & $0.46_{-0.2}^{+0.41}$ & $-0.93_{-0.22}^{+0.16}$ & $-0.83_{-0.51}^{+0.3}$ & $0.03_{-0.001}^{+0.001}$ & $186.0_{-17.0}^{+22.0}$  \\
    J001052.4-002006 & 2.718647 & -0.335062 & SPS & Snellen & 0.94 & $0.09_{-0.007}^{+0.008}$ & $1093_{-167}^{+270}$ & $1.78_{-0.5}^{+1.08}$ & $-0.49_{-0.1}^{+0.12}$ &  &  &   \\
    J001055.6+005951 & 2.7318 & 0.9976 & PS & Snellen & 25.39 & $0.12_{-0.004}^{+0.003}$ & $84_{-3}^{+3}$ & $2.82_{-0.6}^{+0.63}$ & $-0.62_{-0.02}^{+0.02}$ &  &  & \\
    J001130.3+005752 & 2.8764 & 0.9645 & SPS & Retriggered & 34.18 & $0.18_{-0.008}^{+0.01}$ & $2537_{-337}^{+679}$ & $0.53_{-0.03}^{+0.03}$ & $-0.24_{-0.16}^{+0.13}$ & $-0.85_{-0.19}^{+0.14}$ & $0.07_{-0.002}^{+0.002}$ & $240.0_{-18.0}^{+20.0}$  \\
    J001513.8-000337 & 3.8076 & -0.0603 & PS & Retriggered & 30.85 & $0.05_{-0.006}^{+0.016}$ & $471_{-51}^{+45}$ & $0.97_{-0.27}^{+0.48}$ & $-2.29_{-0.44}^{+0.41}$ & $-0.93_{-0.35}^{+0.2}$ & $0.02_{-0.001}^{+0.001}$ & $159.0_{-19.0}^{+18.0}$  \\
    J001611.0-001511 & 4.0459 & -0.2531 & SPS & Snellen & -218.43 & $0.92_{-0.004}^{+0.004}$ & $410_{-10}^{+10}$ & $0.72_{-0.01}^{+0.01}$ & $-0.41_{-0.0}^{+0.0}$ &  &  &  \\
    \hline
    \end{tabular}
    \end{table}
\end{landscape}
\egroup

\section{Additional radio parameters}\label{sec:appxB-radioparams}

We present here the additional radio parameters for this PS sample, including the debiased VI obtained from VAST (Table~\ref{tab:VAST-cat}), and the NSI obtained from MWA-IPS measurements (Table~\ref{tab:IPS-cat}). The tables here contain the first 10 entries, but the full catalogue is available online.

\begin{table*}
    \caption{The variability parameters for our sources as obtained from the VAST Pilot surveys. Column 1 is the IAU name of the source, Column 2 the VAST ID of the source against which each PS object is matched. Columns 3 and 4 are the minimum and maximum signal to noise (SNR) for this source obtained across the VAST Pilot images which contain it. Column 5 is the average flux of the source across all VAST Pilot data. Column 6 is the separation between the position of each object as given in Table~\ref{tab:full-cat} and given by the VAST pipeline. Columns 7 and 8 are the debiased variability index (VI) obtained when using the integrated radio flux density of the source, and the peak radio flux density, respectively. The full version of this table can be found in the online supplementary material.}
    \label{tab:VAST-cat}
\begin{tabular}{lccccccc}
\hline
(1) & (2) & (3) & (4) & (5) & (6) & (7) & (8) \\
IAU Designation & VAST ID & VAST SNR min. & VAST SNR max. & VAST avg. flux & VAST separation & Debiased VI & Debiased VI \\
 & & & & (mJy) & (arcsec) & (int. flux) & (peak flux) \\
\hline
\hline
J035045.7-010848 & 3937280 & 58 & 120 & 20.72 & 0.62 & 0.3008 & 0.2859 \\
J001130.3+005752 & 3587332 & 73 & 631 & 201.28 & 0.80 & 0.2538 & 0.2389 \\
J022213.8-010148 & 3866373 & 39 & 134 & 37.53 & 0.76 & 0.2174 & 0.2056 \\
J035219.8+011421 & 3868128 & 87 & 320 & 68.32 & 0.71 & 0.1958 & 0.1910 \\
J031505.0+011001 & 3748162 & 24 & 128 & 27.17 & 0.57 & 0.1614 & 0.1866 \\
J031357.0+003507 & 4074157 & 36 & 132 & 32.04 & 0.23 & 0.1818 & 0.1720 \\
J224423.5-001712 & 3389059 & 41 & 72 & 17.21 & 0.19 & 0.1444 & 0.1669 \\
J211206.6+001732 & 3870471 & 106 & 202 & 37.35 & 0.41 & 0.1518 & 0.1653 \\
J222744.5+003450 & 4078633 & 118 & 221 & 44.07 & 1.22 & 0.1250 & 0.1484 \\
J031600.9+005854 & 3528910 & 10 & 31 & 8.70 & 0.26 & 0.1641 & 0.1442 \\
\hline
\end{tabular}
\end{table*}
\begin{table*}
    \caption{The IPS parameters of our sources. Most columns are as appear in \citet{Morgan2022}. Again, Column 1 is the IAU designation of each object as it appears in Table~\ref{tab:full-cat}, and Column 2 is the solar elongation of the object during the observations (further from the Sun generally means scintillation will be weaker, even for the most compact sources). Column 3 is the IPS class of an object, whether it is a secure detection (`detected'), a marginal detection, or only an upper limit on scintillation can be calculated. Column 4 is the actual measure of scintillation, NSI, and column 5 its associated error. Columns 6 and 7 are the upper limit on NSI and associated error, where these  are calculated instead of a secure NSI. Column 8 is the  likelihood that the NSI is non-zero, column 9 that the true NSI is not unity. Column 10 is the number of measurements used in calculating the NSI, and column 11 the effective number of observations taking into account the weighting of each observation (this is further explained in \citet{Morgan2022}). Column 12 is the number of 5-$\sigma$ detections of each object in the variability image.}
    \label{tab:IPS-cat}
\begin{tabular}{lllllllllllll}
\hline
(1) & (2) & (3) & (4) & (5) & (6) & (7) & (8) & (9) & (10) & (11) & (12) \\   
IAU Designation & solar elongation & IPS class & NSI$_{\text{fit}}$ & err & UL$_{\text{fit}}$ & err & $\mathcal{L}_0$ & $\mathcal{L}_1$ & $\text{N}_{\text{fit}}$ & $\text{N}_{\text{eff}}$ & $\text{N}_{\text{MOM}}$ \\
& (deg) & & & & & & & & & &\\
\hline
\hline
J000001.6-002209 & 33.09652 & detected & 1.05 & 0.04 &  &  & 3552.89 & 25.17 & 50 & 41.5 & 18 \\
J000452.3-003547 & 32.14316 & detected & 1.10 & 0.07 &  &  & 217.27 & 6.41 & 24 & 21.8 & 2 \\
J001055.6+005951 & 29.859293 & detected & 1.15 & 0.09 &  &  & 166.49 & 9.09 & 16 & 14.6 & 2 \\
J001611.0-001511 & 31.059143 & detected & 1.07 & 0.04 &  &  & 46439.87 & 378.38 & 67 & 51.9 & 46 \\
J002225.4+001456 & 31.17189 & detected & 1.10 & 0.04 &  &  & 439445.20 & 5723.15 & 77 & 56.5 & 68 \\
J004606.6-004341 & 32.35645 & detected & 1.07 & 0.03 &  &  & 322621.70 & 2278.25 & 87 & 63.5 & 73 \\
J004631.3-002023 & 31.81639 & marginal & 0.63 & 0.15 &  &  & 5.75 & 11.91 & 6 & 5.4 & 0 \\
J005205.5+003538 & 33.09957 & marginal & 0.47 & 0.07 &  &  & 14.22 & 163.63 & 35 & 30.4 & 0 \\
J005716.9-002432 & 32.953613 & upper\_limit &  &  & 0.27 & 0.19 & 0.28 & 39.55 & 9 & 8.6& 0 \\
J005905.4+000652 & 31.804417 & detected & 0.96 & 0.03 &  &  & 5145384.50 & 10120.92 & 83 & 59.0 & 83 \\
\hline
\end{tabular}
\end{table*}


\bsp	
\label{lastpage}
\end{document}